\documentclass{optica-article}

\journal{opticajournal} 

\articletype{Research Article}

\usepackage{lineno}
\usepackage{physics}
\usepackage{mathbbol}
\usepackage{comment}

\renewcommand{\ket}[1]{|#1\rangle}
\renewcommand{\bra}[1]{\langle #1|}
\renewcommand{\braket}[2]{\left\langle #1 | #2 \right\rangle}
\renewcommand{\ketbra}[2]{\left| #1 \right\rangle\left\langle #2 \right|}

\begin{document}

\title{On the relevance of weak measurements in dissipative quantum systems}

\author{Lorena Ballesteros Ferraz,\authormark{1,2,*} John Martin,\authormark{2} and Yves Caudano\authormark{1}}

\address{\authormark{1}Research Unit Lasers and Spectroscopies (UR-LLS), naXys \& NISM, University of Namur, Rue de Bruxelles 61, B-5000 Namur, Belgium\\
\authormark{2}Institut de Physique Nucléaire, Atomique et de Spectroscopie, CESAM, University of Liège,
B-4000 Liège, Belgium }

\email{\authormark{*}lorena.ballesteros@unamur.be} 


\begin{abstract*} 
We investigate the impact of dissipation on weak measurements. While weak measurements have been successful in signal amplification, dissipation can compromise their usefulness. More precisely, we show that in systems with non-degenerate eigenstates, weak values always converge to the expectation value of the measured observable as dissipation time tends to infinity, in contrast to  systems with degenerate eigenstates, where the weak values can remain anomalous, i.e., outside the range of eigenvalues of the observable, even in the limit of an infinite dissipation time. In addition, we propose a method for extracting information about the dissipative dynamics of a system using weak values at short dissipation times. Specifically, we explore the amplification of the dissipation rate in a two-level system and the use of weak values to differentiate between Markovian and non-Markovian dissipative dynamics. We also find that weak measurements operating around a weak atom-cavity coupling can probe the atom dissipation through the weak value of non-Hermitian operators within the rotating-wave approximation of the weak interaction.
\end{abstract*}

\section{Introduction}
Weak values are widely used in the context of weak measurements for their amplification capacity \cite{xu2020approaching, hosten2008observation, brunner2010measuring} and the advantageous property of being complex numbers \cite{lundeen2011direct, pati2015measuring}. However, as all quantum systems are open in practice, dissipative effects can affect the amplification qualities of weak values \cite{breuer2002theory}. In this paper, we explore the impact of dissipation on weak measurements. We also look at how weak values, because they are complex and unbounded, can be exploited to extract information on the dissipative dynamics of open quantum systems, such as the dissipation rate or non-markovianity \cite{de2017dynamics}.

The protocol for quantum weak measurements involves four steps and requires a system of interest and a meter (also called ancilla, or pointer), which is used to obtain information about the system \cite{aharonov1988result}. First, the initial state of the system is prepared through pre-selection. In the second step, a weak interaction takes place, which is described by means of a unitary operator that involves both the meter and the system of interest, $\hat{U}=e^{-igt\hat{A}_S\otimes\hat{N}}$, where $\hat{A}_S$ is the system observable to be measured, $\hat{N}$ is a meter operator, $g$ is the interaction strength, and $t$ is the interaction time. Both the interaction strength and interaction time are assumed to be small. This unitary operator is inspired by the von Neumann protocol \cite{mello2014neumann}, but any operator acting on a continuous or discrete state space can act as the meter \cite{svensson2013pedagogical}. The  unitary evolution entangles the system and the meter. In the third step, post-selection is performed to select a particular final state for the system, which involves a projective measurement and filtering. Finally, the meter wave function is read out. Let us consider $\hat{N}$ as the momentum operator. Two shifts appear in the meter wave function: the shift in position representation is proportional to the real part of the weak value, $A_w=\frac{\bra{\psi_f}\hat{A}_S\ket{\psi_i}}{\braket{\psi_f}{\psi_i}}$, where $\ket{\psi_i}$ and $\ket{\psi_f}$ are the pre- and post-selected states respectively, while the shift in momentum representation is proportional to the imaginary part of the weak value. Weak values are unbounded complex numbers. When they diverge, they can be used to amplify tiny signals \cite{jordan2014technical, dixon2009ultrasensitive, harris2017weak}. However, to achieve amplification, the pre- and post-selected states must be almost orthogonal. As a result, the experiment must be repeated multiple times to obtain information about the weak value, because the closer the states are to orthogonality, the lower the probability of post-selection \cite{kofman2012nonperturbative}. However, because of the amplification, the resulting meter shift is much larger than it would have been in the von Neumann protocol without post-selection \cite{svensson2013pedagogical}.

In practice, quantum systems cannot be completely isolated. They always interact with the surrounding environment. The study of these dynamics falls within the scope of open quantum systems theory, which aims to understand how the interaction with the environment affects the system of interest \cite{breuer2002theory, rivas2012open, rotter2015review}. The environment can usually be described in terms of bosonic modes and, in the Born-Markov and secular approximations, the system's dynamics can be modeled by a Lindblad master equation \cite{merkli2022dynamics}, 
\begin{equation}
\label{eq:Lindblad_master_equation}
\dot{\hat{\rho}}=-\frac{i}{\hbar}\left[\hat{H}_S,\hat{\rho}\right]+\sum_i\gamma_i\left(\hat{L}_i\hat{\rho}\hat{L}^{\dagger}_i-\frac{1}{2}\{\hat{L}^{\dagger}_i\hat{L}_i,\hat{\rho}\}\right)\equiv \mathcal{L}\left(\hat\rho\right),
\end{equation}
where $\hat{L}_i$ are a set of jump operators and $\gamma_i$ are dissipation rates. In the following, to describe the time evolution over a time $t$ of the operator $\hat{\rho}$ governed by the Lindblad master equation, we will use the superoperator notation $e^{\mathcal{L}t}\hat{\rho}$. Equation \eqref{eq:Lindblad_master_equation} consists of two terms: the first term represents the unitary evolution of the density operator according to von Neumann's equation, while the second term, also noted $\mathcal{D}\left(\hat\rho\right)$, accounts for the non-unitary dynamics resulting from dissipation, decoherence and dephasing. The dissipator $\mathcal{D}$ involves dissipation rates $\gamma_i$, one for each dissipation channel present \cite{breuer2002theory}.

Wiseman introduced the concept of weak values in dissipative systems in the context of homodyne measurements \cite{wiseman2009quantum}. Since then, a few studies investigated weak measurements in open quantum systems. Two studies focused on the detrimental effects of decoherence on weak values, in particular, on the possibility that decoherence limits the sensitivity of sensors based on weak value amplification \cite{knee2013quantum, shikano2009weak}. Other research  explored how Markovian environments prevent weak values from exhibiting anomalous properties and how quickly this process occurs \cite{ban2013weak, abe2015decoherence, shikano2009weak, ban2017weak} (an anomalous weak value has a value different from any average of the observable). Studies also sought to identify the optimal combination of a reservoir and a quantum system that minimizes the detrimental effects of dissipation on weak values at any given time. Non-Markovian environments appear to degrade the anomalous properties of weak values more slowly \cite{abe2016decoherence}.

Our work expands upon previous studies by exploring the general limit of weak values at large dissipation times. Furthermore, we study the case of systems with degenerate ground states in the context of weak measurements and show that these systems can preserve the anomalous behaviour of weak values even in the limit of infinite dissipation times.

In addition, we leverage the amplification properties of weak values at short dissipation times and small dissipation rates to extract valuable information about the dissipation process. This approach serves as a valuable supplementary experiment, shedding light on the significance of incorporating dissipation rates in the modeling process of the main experimental setup, particularly when there is limited time available for dissipation. After a brief dissipation period, it enables an effective measurement of an amplified interaction rate, facilitating a comprehensive evaluation of its implications in the original experiment. 
As after a short dissipation, one could measure an amplified interaction rate. Specifically, we focus on weak measurements where dissipation occurs after the weak interaction and before post-selection. The sequence involves pre-selecting the system, applying a general unitary operator $\hat{U}=e^{-igt\hat{A}_S\otimes\hat{N}}$, allowing for dissipative dynamics during a time $\tau$, and finally performing post-selection. This scheme is present in any experimental setup with a time delay between the system-meter interaction and post-selection. We assume that the duration of the weak interaction is sufficiently short, so that any dissipation during this period is negligible.

We finally consider how our protocol would perform in a specific setup involving a two-level atom as the system and a single-mode cavity field as the meter. The weak measurement relies on a small interaction between the atom and the cavity field during the atom's short transit through the cavity. Subsequently, the atom undergoes dissipation through its interaction with the quantized radiation field of free space or another (leaky) cavity through which it passes. Post-selection is then performed and the real and imaginary parts of the weak value are measured by reading out the $\hat{Q}$ or $\hat{P}$ quadrature of the cavity field. The dissipation rate can be inferred by measuring the meter state with the benefits of weak value amplification. Additionally, if the dissipation process is non-Markovian, under certain circumstances, we show that it can be distinguished from a Markovian one based on the amplified weak value.

\section{General weak measurements with dissipation}\label{sec:generalTheory}
\begin{figure}[t]
 \centering
\includegraphics[width=1\textwidth]{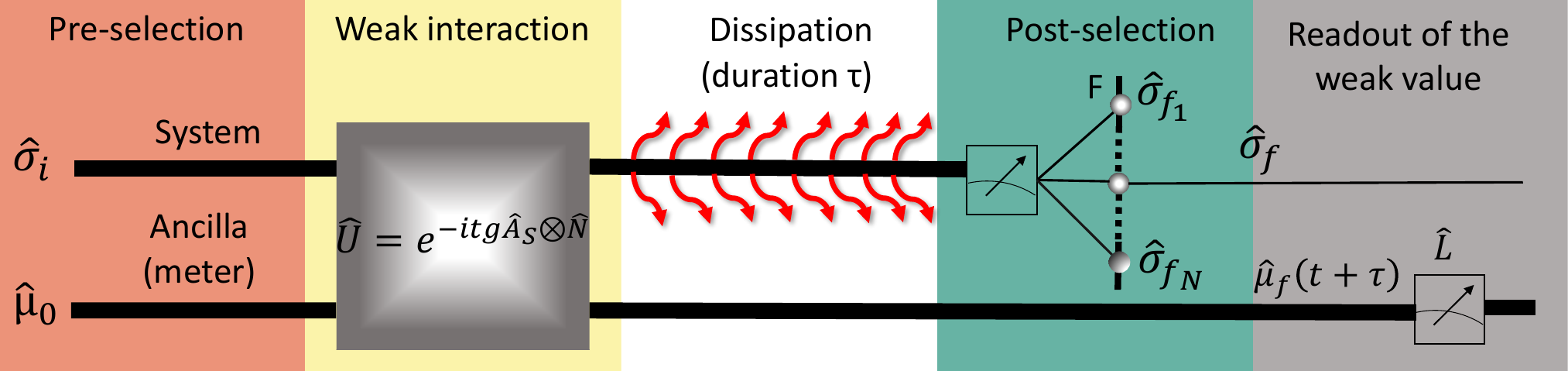}
\caption{Overview of the steps involved in the dissipative weak measurement protocol. The process begins with pre-selection of the system, followed by the implementation of a weak measurement described through the unitary operator $\hat{U}$. After the weak interaction, the system undergoes dissipation for a duration $\tau$ before a post-selection. Finally, the meter is readout to extract information on the weak value.\label{fig:scheme_dissipative_weak_measurements}}
\end{figure}

In this section, we consider the general theory of weak measurements, following the procedure of \cite{svensson2013pedagogical}, but adding a dissipative evolution. The dissipation occurs during the time delay between the weak system--meter interaction and the post-selection. The protocol consists of five steps: system pre-selection, weak interaction, dissipative dynamics, post-selection on the system, and meter readout, as illustrated in Fig.~\ref{fig:scheme_dissipative_weak_measurements}.

Consider that the initial state of the system is described by the density operator $\hat{\sigma}_i$ and that of the meter by $\hat{\mu}_0$. The tensor product of the states of the system and the meter provides the state of the full composite system, resulting in 
\begin{equation}
\label{eq:full_system_state}
\hat{\rho}_0=\hat{\sigma}_i\otimes\hat{\mu}_0. 
\end{equation}
The full system evolution depends on three components: the system Hamiltonian $\hat{H}_S$, the meter (or ancilla) Hamiltonian $\hat{H}_A$, and the interaction Hamiltonian
\begin{equation}
\label{eq:interaction_hamiltonian}
\hat{H}_{\mathrm{int}}=\hbar g\hat{A}_S\otimes\hat{N},
\end{equation}
where $\hat{A}_S$ is the observable of interest acting in the system Hilbert space, $\hat{N}$ is an operator acting in the meter Hilbert space, and $g$ is the interaction strength. 
In the interaction picture with respect to $\hat{H}_0=\hat{H}_S\otimes\hat{\mathbb{1}}+\hat{\mathbb{1}}\otimes\hat{H}_A$, the density operator of the composite system evolves as
\begin{equation}\label{rhoILindblad}
\frac{d \hat{\rho}_I\left(t\right)}{d t}=-\frac{i}{\hbar}\left[\hat{V}\left(t\right),\hat{\rho}_I\left(t\right)\right],
\end{equation} 
where the interaction Hamiltonian and the global density operator are given in the interaction picture by 
\begin{equation}
\hat{V}\left(t\right)=e^{\frac{i\hat{H}_0t}{\hbar}}\hat{H}_{\mathrm{int}}\,e^{\frac{-i\hat{H}_0t}{\hbar}}, \hspace{1 cm} \hat{\rho}_I\left(t\right)=e^{\frac{i\hat{H}_0t}{\hbar}}\hat{\rho}_0\,e^{\frac{-i\hat{H}_0t}{\hbar}}.
\end{equation}
The solution of Eq.~\eqref{rhoILindblad} reads 
\begin{equation}\label{rhoIsol}
\hat{\rho}_I\left(t\right)=\overrightarrow{\mathscr{T}}\text{exp}\left[-\frac{i}{\hbar}\int_0^t\hat{V}\left(t'\right)dt'\right]\hat{\rho}_I\left(0\right)\text{exp}\left[\frac{i}{\hbar}\int_0^t\hat{V}\left(t'\right)dt'\right]\overleftarrow{\mathscr{T}},
\end{equation}   
where $\mathscr{T}$ is the time ordering operator. Given the small interaction strength $g$ and the fact that $\hat{V}(t)\propto g$, we can expand the exponential in Eq.~\eqref{rhoIsol} in a Taylor series to first order in the interaction strength. As a result, we can express the density operator at time $t$ as follows
\begin{eqnarray}
\label{eq:evolution_rho_interaction_picture_after_short_time}
\hat{\rho}_I\left(t\right)&\approx&\left[\hat{\mathbb{1}}-\frac{i}{\hbar}\int_0^t\hat{V}\left(t'\right)dt'\right]\hat{\rho}_0\left[\hat{\mathbb{1}}+\frac{i}{\hbar}\int_0^t\hat{V}\left(t'\right)dt'\right] \\ \nonumber
&\approx& \hat{\rho}_0+\frac{i}{\hbar}\hat{\rho}_0\int_0^t \hat{V}\left(t'\right)dt'-\frac{i}{\hbar}\int_0^t  \hat{V}\left(t'\right)dt'\hat{\rho}_0\\ \nonumber
&=&\hat{\rho}_0+\frac{i}{\hbar}\left[\hat{\rho}_0, \int_0^t  \hat{V}\left(t'\right)dt'\right].
\end{eqnarray}
We assume that the weak interaction duration $t$ is sufficiently short, so that the interaction Hamiltonian $\hat{V}(t)$ does not evolve much from $\hat{H}_{\mathrm{int}}$ under the free evolution due to $\hat{H}_0$. In these circumstances, we can make a series development of $\hat{V}(t^\prime)$ around $t/2$ to first order:
\begin{equation}\label{eq:interactionHamiltonianApprox}
    \hat{V}\left(t^\prime\right)\approx \hat{V}\left(t/2\right)+\dot{\hat{V}}(t/2)\left(t^\prime-t/2\right)
\end{equation}
with $\dot{\hat{V}}(t/2)=\frac{d\hat{V}}{dt'}\big|_{t'=t/2}$.
Then, the density operator is given, at first order in time, by
\begin{eqnarray}
\label{eq:evolution_rho_interaction_picture_after_short_time_simplified}
\hat{\rho}_I\left(t\right)&\approx&\hat{\rho}_0+\frac{i}{\hbar}\left[\hat{\rho}_0, \int_0^t dt'\left(
\hat{V}\left(t/2\right)+\dot{\hat{V}}(t/2)\left(t^\prime-t/2\right)
\right)\right] \\ \nonumber
&=& \hat{\rho}_0+\frac{i}{\hbar}t\left[\hat{\rho}_0, \hat{V}\left(t/2\right)\right],
\end{eqnarray}
since the integral of the second term is exactly 0. The first  contribution neglected in (\ref{eq:evolution_rho_interaction_picture_after_short_time_simplified}) due to the approximation (\ref{eq:interactionHamiltonianApprox}) is of the order $\left(\frac{i}{2\hbar}\right)^2 \frac{1}{3!}t^3\left[\hat{H}_0,\left[\hat{H}_0,\hat{V}\left(t/2\right)\right]\right] $.

After the unitary evolution of the composite system, the system $S$ follows a dissipative dynamics during a time interval $\tau$, while it is assumed that the meter does not undergo any dissipative process. The total density operator $\hat{\rho}_I\left(t+\tau\right)$ is given, after the dissipative evolution, by
\begin{equation}
\label{eq:evolution_rho_interaction_picture_after_dissipation}
\hat{\rho}_I\left(t+\tau\right)=\left(e^{\mathcal{D}\tau}\otimes\hat{\mathbb{1}}\right)\hat{\rho}_I\left(t\right)\approx \left(e^{\mathcal{D}\tau}\otimes\hat{\mathbb{1}}\right)\left(\hat{\rho}_0+\frac{i}{\hbar}t\left[\hat{\rho}_0, \hat{V}\left(t/2\right)\right] \right),
\end{equation}
where the dissipator $\mathcal{D}$ has replaced the full Linbladian $\mathcal{L}$ because we are working in the interaction picture. Indeed, in the interaction picture, the evolution defined in (\ref{eq:Lindblad_master_equation}) becomes $\dot{\hat\rho}_I=\mathcal{D}\left(\rho_I\right)$. Note that the dissipator remains identical to the one defined in the Schrödinger representation, as a result of the commutation rules obeyed by the jump operators $\hat{L}_i$ with the system Hamiltonian $\hat{H}_S$ \cite{Manzano2020}. After the dissipation time $\tau$, the system is post-selected to the final state $\hat{\sigma}_f$. Post-selection is carried out by performing a projective measurement and filtering the information relevant to the state $\hat{\sigma}_f$. At this point, the meter state is, in the interaction picture,
\begin{equation}
\hat{\mu}_{fI}\left(t+\tau\right)=\frac{\text{Tr}_S\left[\left(\hat{\sigma}_{fI}\left(t+\tau\right)\otimes\hat{\mathbb{1}}\right)\left(e^{\mathcal{D}\tau}\otimes\hat{\mathbb{1}}\right)\hat{\rho}_I\left(t\right)\right]}{\text{Tr}\left[\left(\hat{\sigma}_{fI}\left(t+\tau\right)\otimes\hat{\mathbb{1}}\right)\left(e^{\mathcal{D}\tau}\otimes\hat{\mathbb{1}}\right)\hat{\rho}_I\left(t\right)\right]}, 
\label{eq:density_matrix_meter_after_post-selection}
\end{equation}
where $\text{Tr}_S$ is the partial trace over the system degrees of freedom, and the post-selection operator in the interaction picture depends on the post-selection time $t+\tau$ as
\begin{equation}
\label{eq:post_selected_state_interaction_picture}
\hat{\sigma}_{fI}\left( t+ \tau\right)=e^{\frac{i}{\hbar}\hat{H}_S\left(t+\tau\right)}\hat{\sigma}_{f} e^{-\frac{i}{\hbar}\hat{H}_S\left(t+\tau\right)}. 
\end{equation}
In the remainder of this section, we will drop the explicit time dependence of $\hat{\sigma}_{fI}\left( t+ \tau\right)$ and note it $\hat{\sigma}_{fI}$ for simplicity. 

The expression in the denominator of Eq. (\ref{eq:density_matrix_meter_after_post-selection}) corresponds to the probability of obtaining the post-selected state $\hat{\sigma}_f$, given the state of the composite system $\hat{\rho}_I\left(t+\tau\right)$. The denominator is equal to the trace of the numerator, ensuring that the trace of the meter state is equal to 1. Using $\hat{H}_{\mathrm{int}}=\hbar g\hat{A}_S\otimes\hat{N}$, the denominator of Eq.~(\ref{eq:density_matrix_meter_after_post-selection}) becomes, to first order in $g t$,
\begin{eqnarray}
& &\mkern-72mu\text{Tr}\left[\left(\hat{\sigma}_{fI}\otimes\hat{\mathbb{1}}\right) \left(e^{\mathcal{D}\tau}\otimes\hat{\mathbb{1}}\right)\hat{\rho}_I\left(t\right)\right]\nonumber\\
&\approx&\text{Tr}\left[\hat{\sigma}_{fI}\ e^{\mathcal{D}\tau}\left(\hat{\sigma}_i\right)\right]+2gt\,\text{Im}\left(\text{Tr}\left[\hat{\sigma}_{fI}\ e^{\mathcal{D}\tau}\left(\hat{A}_{SI}\!\left(t/2\right)\hat{\sigma}_i\right)\right]\text{Tr}\left[\hat{\mu}_0\hat{N}_{I}\!\left(t/2\right)\right]\right) \nonumber \\
&=&\text{Tr}\left[\hat{\sigma}_{fI}\ e^{\mathcal{D}\tau}\left(\hat{\sigma}_i\right)\right] \left( 1+2 g t \Im \left[A_{S,w}\left(\tau\right)\right] \langle\hat{N}_I\!\left(t/2\right)\rangle_{0}\right),\label{eq:probability_of_post-selection}
\end{eqnarray}
where we have defined the weak value with dissipation, denoted $A_{S,w}\left(\tau\right)$, as 
\begin{equation}\label{eq:weak_value_with_dissipation}
A_{S,w}\left(\tau\right)=\frac{\text{Tr}\left[\hat{\sigma}_{fI}\ e^{\mathcal{D}\tau}\left(\hat{A}_{SI}\!\left(t/2\right)\hat{\sigma}_i\right)\right]}{\text{Tr}\left[\hat{\sigma}_{fI}\ e^{\mathcal{D}\tau}\left(\hat{\sigma}_i\right)\right]},
\end{equation}
and $\langle\hat{N}_I\!\left(t/2\right)\rangle_{0}=\text{Tr}\left[\hat{\mu}_0\hat{N}_I\!\left(t/2\right)\right]$ corresponds to the expectation value of the operator $\hat{N}_I\!\left(t/2\right)$ in the initial meter state $\hat{\mu}_0$. Our definition \eqref{eq:weak_value_with_dissipation} is very general and coincides with Wiseman's definition in the special case of homodyne measurements in cavity QED~\cite{wiseman2002weak}. We stress that the weak value (\ref{eq:weak_value_with_dissipation}) has been calculated in the interaction picture and that the system and meter observables $\hat{A}_S$ and $\hat{N}$ appear thus as $A_{SI}\!\left(t/2\right)$ and $N_{I}\!\left(t/2\right)$, respectively. Their time dependence at $t/2$ results from the approximation (\ref{eq:interactionHamiltonianApprox}). The operator $\hat{\sigma}_{fI}$ corresponds to the post-selected state $\hat{\sigma}_{f}$  at time $t +\tau$ in the interaction picture, as expressed in Eq.~(\ref{eq:post_selected_state_interaction_picture}). If we revert to the Schrödinger picture, we see that $t/2$ corresponds to the effective time of the pre-selection and the post-selection for an instantaneous weak interaction that is symmetric with respect to pre- and post-selection. Advantageously, by setting the post-selection operator as a constant in the interaction picture, namely $\hat\sigma_{fI}\left(t+\tau\right)=\hat\sigma_{fI}\left(t/2\right)$ (the constant is defined with respect to the effective pre- and post-selection time), we can suppress in practice the effects of the system free evolution during time $\tau+t/2$ on the weak value, so that we can focus specifically on the repercussions of dissipation, as analyzed later on in this paper.

Similarly, for the numerator of Eq.~(\ref{eq:density_matrix_meter_after_post-selection}), we obtain 
\begin{eqnarray}
&&\mkern-45mu \text{Tr}_S\left[\left(\hat{\sigma}_{fI}\otimes\hat{\mathbb{1}}\right)\left(e^{\mathcal{D}\tau}\otimes\hat{\mathbb{1}}\right)\hat{\rho}_I\left(t\right)\right] \nonumber\\ \nonumber
&\approx&\text{Tr}\left[\hat{\sigma}_{fI}\ e^{\mathcal{D}\tau}\left(\hat{\sigma}_i\right)\right]\hat{\mu}_0+igt\left\{\text{Tr}\left[\hat{\sigma}_{fI}\ e^{\mathcal{D}\tau}\left(\hat{\sigma}_i\hat{A}_{SI}\right)\right]\hat{\mu}_0\hat{N}_{I}-\text{Tr}\left[\hat{\sigma}_{fI}\ e^{\mathcal{D}\tau}\left(\hat{A}_{SI}\hat{\sigma}_i\right)\right]\hat{N}_{I}\hat{\mu}_0\right\}\\ 
&=&\text{Tr}\left[\hat{\sigma}_{fI}\ e^{\mathcal{D}\tau}\left(\hat{\sigma}_i\right)\right]\left[\hat{\mu}_0+igt\left(\overline{A_{S,w}\left(\tau\right)}\hat{\mu}_0\hat{N}_{I}-A_{S,w}\left(\tau\right)\hat{N}_{I}\hat{\mu}_0\right)\right],\label{eq:meterStateNum}
\end{eqnarray}
where the interaction picture operators $\hat{A}_{SI}$ and $\hat{N}_{I}$ should be evaluated at time $t/2$, and where we used the property $\left[e^{\mathcal{D}\tau}\left(\hat{O}\right)\right]^{\dagger}=e^{\mathcal{D}\tau}\left(\hat{O}^{\dagger}\right)$, valid for an arbitrary operator $\hat{O}$.

By combining the denominator (\ref{eq:probability_of_post-selection}) and the numerator (\ref{eq:meterStateNum}),
the final meter density matrix is, at first order in $g t$,
\begin{equation}
\label{eq:final_state_meter}
\hat{\mu}_{fI}\left(t+\tau\right)=\frac{\hat{\mu}_0+igt\left(\overline{A_{S,w}\left(\tau\right)}\hat{\mu}_0\hat{N}_I\!\left(t/2\right)-A_{S,w}\left(\tau\right)\hat{N}_I\!\left(t/2\right)\hat{\mu}_0\right)}{1+2 g t \Im \left[A_{S,w}\left(\tau\right)\right] \langle\hat{N}_I\!\left(t/2\right)\rangle_0}.
\end{equation}
Up to a normalization constant, the final meter state can be expressed as the initial meter state $\hat\mu_0$ plus a term that depends on the pre-selected state, the observable, and the post-selected state. While this new term may seem small in principle, as it depends on a small parameter $gt$, there are cases where the weak value $A_{S,w}\left(\tau\right)$ becomes large, specifically when its denominator is close to zero due to the near-orthogonality of the pre- and post-selected states, resulting in a significant contribution to the meter state. 

To actually perform the weak measurement, one should measure the expectation value of a meter observable, $\hat{L}$. The measurement result will depend on the weak value that appears in the meter final state, as described by Eq.~(\ref{eq:final_state_meter}). In the interaction picture, $\hat{L}$ is given by
\begin{equation}
\label{eq:L_interaction_picture}
\hat{L}_I\left(t+\tau\right)=e^{\frac{i}{\hbar}\hat{H}_A (t+\tau)}\hat{L}\, e^{-\frac{i}{\hbar}\hat{H}_A (t+\tau)},
\end{equation}
while its expectation value takes the form 
\begin{equation}
\label{eq:expectation_value_L_dissipation_WithDenom}
\langle\hat{L}\rangle_f=\text{Tr}\left(\hat{L}_I\left(t+\tau\right)\hat{\mu}_{fI}\right)=\frac{\langle\hat{L}_I\!\left(t+\tau\right)\rangle_0+2gt\,\text{Im}\left[A_{S,w}\left(\tau\right)\langle\hat{L}_I\!\left(t+\tau\right)\hat{N}_I\!\left(t/2\right)\rangle_0\right]}{1+2 g t \Im \left[A_{S,w}\left(\tau\right)\right] \langle\hat{N}_I\!\left(t/2\right)\rangle_0},
\end{equation}
where a procedure similar to \cite{svensson2013pedagogical} was followed to obtain the numerator expression. All expectation values are computed with respect to the initial meter state $\hat\mu_0$.

From now on, let us assume that the expectation value of the meter operator $\hat{N}_I\left(t/2\right)$ in the meter initial state is zero, that is, $\langle\hat{N}_I\!\left(t/2\right)\rangle_0=0$. If this is not the case, we could possibly redefine and translate the meter operator $\hat{N}$ to satisfy $\langle\hat{N}_I\!\left(t/2\right)\rangle_0=0$. When the meter free evolution is fully negligible during the short interaction time $t$, this assumption states that the expectation value of the meter observable is zero in the meter initial state, a natural calibration requirement imparted to the meter initial state. In the Schrödinger picture, the assumption requires the expectation value of the meter observable $\hat{N}$ to be zero in the state $\hat\mu_0\!\left(t/2\right)$ resulting from the free evolution for a duration $t/2$ of the meter initial state $\hat\mu_0$. In these conditions, the meter expectation value at the end of the measurement is simply the numerator of (\ref{eq:expectation_value_L_dissipation_WithDenom}), namely
\begin{equation}
\label{eq:expectation_value_L_dissipation}
\langle\hat{L}\rangle_f=\text{Tr}\left(\hat{L}_I\left(t+\tau\right)\hat{\mu}_{fI}\right)=\langle\hat{L}_I\!\left(t+\tau\right)\rangle_0+2gt\,\text{Im}\left[A_{S,w}\left(\tau\right)\langle\hat{L}_I\!\left(t+\tau\right)\hat{N}_I\!\left(t/2\right)\rangle_0\right].
\end{equation}
In addition, if we select an initial meter density operator that commutes with the meter Hamiltonian, i.e.\ $\left[\hat{\mu}_0,\hat{H}_A\right]=0$, the final meter average simplifies to 
\begin{equation}
\label{eq:expectation_value_L_dissipation_without_commutator}
\langle\hat{L}\rangle_f=\langle\hat{L}\rangle_0+2 g t \,\text{Im}\left[A_{S,w}\left(\tau\right)\langle\hat{L}_I\!\left(t/2+\tau\right)\hat{N}\rangle_0\right],
\end{equation}
where only the operator $\hat{L}_I$ remains in the interaction picture, evaluated at the effective time $t/2+\tau$. The latter reflects the natural
evolution under $\hat{H}_A$ of the meter state perturbation due to the weak interaction. For short dissipation times $\tau$, this time dependence could be neglected whenever $\frac{i}{\hbar}\left(t/2+\tau\right)\langle\left[\hat{H}_A,\hat{L}\right]\hat{N}\rangle_0\ll\langle\hat{L}\hat{N}\rangle_0$.

The shift in the expectation value of a general meter observable, $\hat{L}$, as given by Eq.~(\ref{eq:expectation_value_L_dissipation}), depends on both the real and imaginary parts of the weak value. It is enlightening to reformulate it the following way:
\begin{equation}
\label{eq:expectation_value_L_RealCommutator-ImaginaryAntiCommutator}
\langle\hat{L}\rangle_f=\langle\hat{L}_I\rangle_0 - i g  t\,\text{Re}A_{S,w}\left(\tau\right)\langle\left[\hat{L}_I,\hat{N}_I\right]\rangle_0 + g  t\,\text{Im}A_{S,w}\left(\tau\right)\langle\left\{\hat{L}_I,\hat{N}_I\right\}\rangle_0,
\end{equation}
where the interaction operators should be evaluated at the appropriate times $\hat{L}_I\!\left(t+\tau\right)$ and $\hat{N}_I\!\left(t/2\right)$. This expression makes the real and imaginary parts of the weak value appear explicitly.
We can choose $\hat{L}$ such that $\langle \hat{L}_I\!\left(t+\tau\right)\rangle_0=0$, to ensure that the expectation value of $\hat{L}$ is now directly proportional to the terms which depend on the weak value. To separate the shifts due to the real and imaginary components, we can choose specific meter observables. For instance, when the meter observable at time $t+\tau$ is equal to the pointer at time $t/2$, i.e.\ $\hat{L}_I\!\left(t+\tau\right)=\hat{N}_I\!\left(t/2\right)$, the expectation value of the meter observable is proportional to the imaginary part of the weak value. Indeed, in this case, the expectation value (\ref{eq:expectation_value_L_RealCommutator-ImaginaryAntiCommutator}) reads
\begin{equation}
\label{eq:expectation_value_N}
\langle\hat{N}\rangle_f= 2 g t\langle\hat{N}_I^2\!\left(t/2\right)\rangle_0\,\text{Im} A_{S,w}\left(\tau\right),
\end{equation}
where $\langle\hat{N}^2_I\!\left(t/2\right)\rangle_0=\text{Tr}\left[\hat{\mu}_0\hat{N}^2_I\!\left(t/2\right)\right]=\Delta^2 \hat{N}_I\!\left(t/2\right)\neq 0$, considering that $\langle\hat{N}_I\!\left(t/2\right)\rangle_0=0$. On the other hand, when the meter observable, $\hat{L}=\hat{M}$, at time $t+\tau$ is the canonical conjugate of the pointer at time $t/2$, such that $\left[\hat{M}_I\!\left(t+\tau\right),\hat{N}_I\!\left(t/2\right)\right]=i\hbar\hat{\mathbb{1}}$, its expectation value has a term proportional to the real part of the weak value. More specifically, we have 
\begin{equation}
\label{eq:expectation_value_M}
\langle\hat{M}\rangle_f=\left[\hbar g t \ \text{Re} A_{S,w}\left(\tau\right)+ g t \ \langle\left\{\hat{M}_I\!\left(t+\tau\right),\hat{N}_I\!\left(t/2\right)\right\}\rangle_0\ \text{Im} A_{S,w}\left(\tau\right)\right],
\end{equation}
assuming that $\langle\hat{M}_I\!\left(t+\tau\right)\rangle_0=0$. It is sometimes possible to choose the observables and the initial state in such a way that we also have $\langle\left\{\hat{M}_I\!\left(t+\tau\right),\hat{N}_I\!\left(t/2\right)\right\}\rangle_0=0$. This then leads to the expectation value of $\hat{M}$ being directly proportional to the real part of the weak value. Using $\hat{N}$ and $\hat{M}$ as meter observables enables separating the real and imaginary components of the weak value, which is often desirable in many experimental setups. 
In addition, they enable describing the argument of the weak value \cite{mc2016, mc2017}, here as a function of the dissipation time $\tau$, in the meter phase space \cite{lbf2022}. In section \ref{sec:RabiModel}, we will show how Eq.~(\ref{eq:expectation_value_L_dissipation}) connects directly the meter measurement to the modulus and argument of the weak value. We remind the reader that equations (\ref{eq:expectation_value_L_dissipation}--\ref{eq:expectation_value_M}) were obtained under the assumptions that the expectation value $\langle\hat{N}_I\left(t/2\right)\rangle_0=0$. If this is not the case, the denominator present in (\ref{eq:expectation_value_L_dissipation_WithDenom}) should be kept. In particular, the general expression for (\ref{eq:expectation_value_L_RealCommutator-ImaginaryAntiCommutator}) is
\begin{equation}
\label{eq:expectation_value_L_RealCommutator-ImaginaryAntiCommutator_WithDenominatorAdded}
\langle\hat{L}\rangle_f=\frac{\langle\hat{L}_I\rangle_0 - i g  t\,\text{Re}A_{S,w}\left(\tau\right)\langle\left[\hat{L}_I,\hat{N}_I\right]\rangle_0 + g  t\,\text{Im}A_{S,w}\left(\tau\right)\langle\left\{\hat{L}_I,\hat{N}_I\right\}\rangle_0}{1+2 g t \Im \left[A_{S,w}\left(\tau\right)\right] \langle\hat{N}_I\rangle_0},
\end{equation}
with the appropriate time dependence $\hat{L}_I\!\left(t+\tau\right)$ and $\hat{N}_I\!\left(t/2\right)$.

The main results of this section, namely Eqs.~(\ref{eq:weak_value_with_dissipation}), (\ref{eq:expectation_value_L_dissipation}) and \eqref{eq:expectation_value_L_RealCommutator-ImaginaryAntiCommutator}, as well as (\ref{eq:expectation_value_L_RealCommutator-ImaginaryAntiCommutator_WithDenominatorAdded}), describe the consequences on weak measurements of dissipation occurring after the weak interaction, before post-selection. These expressions ensue from the weak interaction approximation (\ref{eq:interactionHamiltonianApprox}--
\ref{eq:evolution_rho_interaction_picture_after_short_time_simplified}), that was used in (\ref{eq:probability_of_post-selection}) and (\ref{eq:meterStateNum}) to evaluate the meter reduced density matrix (\ref{eq:density_matrix_meter_after_post-selection}) to first order in $g t$. By using expressions in the interaction picture, we also include the full treatment of the free Hamiltonian evolution of both the system and the meter, which is required for describing long dissipation times $\tau$.
While there could also be dissipation during the time delay between pre-selection and the system--meter interaction, this would only alter the initial system state from $\hat{\sigma}_i\left(-T\right)$ to $\hat{\sigma}_i\left(0\right)$, where $\hat{\sigma}_i\left(-T\right)$ is the density operator produced by the pre-selection procedure at time $-T$ (either as characterized experimentally or as defined theoretically) and where $T$ is the time delay between pre-selection and the application of the unitary operator at time 0. In that case, in the definition (\ref{eq:weak_value_with_dissipation}) of the weak value with dissipation, we should simply use $\hat{\sigma}_i\left(0\right)$ for the effectively pre-selected density operator: $\hat{\sigma}_i=\hat{\sigma}_i\left(0\right)=e^{\mathcal{L}T}\left(\hat{\sigma}_i\left(-T\right)\right)=e^{\mathcal{D}T}\left(\hat{\sigma}_{iI}\left(-T\right)\right)$, with the last equality expressed in the interaction picture. The effect of dissipation before the weak interaction is thus simply that the effective initial state $\hat{\sigma}_i\left(0\right)$ may differ from the desired initial state. As a result, it is generally possible to modify the post-selected state in order to partially preserve the amplification capabilities of the weak value (if this is the objective), provided that the evolution of the system is well-known and that the actual initial state $\hat{\sigma}_i\left(-T\right)$ is not completely mixed. In general, Eq. (\ref{eq:expectation_value_L_dissipation}) does not account for dissipation occurring during the weak interaction, unless we assume that it is negligible because of the short duration $t$ of the  weak interaction. 

As an illustration of the impact of dissipation before the weak interaction, let us consider the case of dissipation occurring only before the weak interaction (no dissipation in between the weak interaction and post-selection). In that situation, the weak value would be
\begin{equation}
\widetilde{A_{S,w}}\left(T\right)=\frac{\text{Tr}\left[\hat{\sigma}_{fI}\!\left(t+\tau\right)\hat{A}_{SI}\!\left(t/2\right) e^{\mathcal{D}T}\left(\hat{\sigma}_{iI}\!\left(-T\right)\right)\right]}{\text{Tr}\left[\hat{\sigma}_{fI}\!\left(t+\tau\right) e^{\mathcal{D}T}\left(\hat{\sigma}_{iI}\!\left(-T\right)\right)\right]},
\end{equation}
where $T$ is the dissipation duration. If the system state at time $-T$ is not completely mixed, the pre-selected state evolution under dissipation can be taken into account by modifying the post-selected state, in order to preserve some amplification, even at a very large dissipation time $T$. In other words, we can often choose a post-selected state that is sufficiently orthogonal to the pre-selected state after dissipation, $\hat\sigma_i\left(0\right)$. As an illustration, if the dissipation time is infinite and the system possesses a single non-degenerate ground state $\ket{g}$, considering a pure post-selected state $\ket{\psi_f}$, and ignoring the effects of the system free evolution, the weak value is
\begin{equation}
\lim_{T\rightarrow\infty}\widetilde{A_ {S,w}}\left(T\right)=\frac{\bra{\psi_f}\hat{A}\ket{g}}{\braket{\psi_f}{g}}.
\end{equation}
By choosing a post-selected state that is almost orthogonal to $\ket{g}$, we can find amplification even at infinite dissipation time. We will show in the next section that this is not the case when dissipation takes place after the weak interaction.

The consequences on the weak value of having dissipation before or after the weak interaction are completely different. Having dissipation before the weak interaction simply alters the initial state. However, having dissipation after the weak interaction destroys the coherences of the system, partially or completely, as we will show in the next section. Consequently, both types of dissipation should be studied separately. Furthermore, all the results of this paper can be extended to the case in which there is dissipation before the weak interaction by changing the initial density operator, $\hat{\sigma}_i\left(-T\right)$ by an initial density operator after dissipation $\hat{\sigma}_i\left(0\right)$, just before the weak interaction.

\section{Weak value evolution}\label{sec:WVevol}
In this section, we investigate the properties of the weak value of an arbitrary operator in the context of long dissipation times. Previous studies have suggested that dissipation can destroy the anomalous properties of the weak value \cite{knee2013quantum,  ban2013weak, abe2015decoherence, shikano2009weak, ban2017weak, abe2016decoherence}, which means that weak values converge to values within the  range of the eigenvalues of the observable in the presence of dissipation. Our findings reveal that, in non-degenerate systems, the weak value approaches the expectation value of the operator as the dissipation time tends to infinity. Furthermore, we demonstrate that the anomalous properties of the weak value can persist at infinite dissipation time in systems with degenerate ground states. This suggests that dissipation does not always necessarily result in the loss of the amplification effect that the weak value can provide.

\subsection{Non-degenerate ground state}
Let us assume that the system under study possesses only one ground state, denoted by $\ket{g}$. In this case, the dissipative evolution invariably destroys anomalous properties of weak values at very long times, regardless of the system's markovianity. For simplicity of calculation, and without loss of generality, let us consider a two-level system. The excited state of the system is denoted by $\ket{e}$, and we assume that the pre- and post-selected states, $\ket{\psi_i}$ and $\ket{\psi_f\left(\tau\right)}$, are both pure states. We choose a post-selected state that depends on the dissipation duration $\tau$ in such a way that the post-selected state is constant in the interaction representation, noted $\ket{\psi_{fI}}$. This allows us to focus on analyzing the consequences of dissipation without observing effects of the free Hamiltonian evolution of the system during time $\tau$. In practice, this is equivalent to ensuring that the effectively post-selected state at time $t/2$ does not depend on the dissipation duration. In this scenario, the weak value with dissipation can be expressed as follows
\begin{eqnarray}
&&\mkern-36mu A_{S,w}\left(\tau\right)=\frac{\text{Tr}\left[\ket{\psi_{fI}}\bra{\psi_{fI}}e^{\mathcal{D}\tau}\left(\hat{A}_{SI}\ket{\psi_i}\bra{\psi_i}\right)\right]}{\text{Tr}\left[\ket{\psi_{fI}}\bra{\psi_{fI}}e^{\mathcal{D}\tau}\left(\ket{\psi_i}\bra{\psi_i}\right)\right]}\\ \nonumber
&&=\frac{\text{Tr}\left[\ket{\psi_{fI}}\bra{\psi_{fI}}\left(D_{ee}\left(\tau\right)\ket{e}\bra{e}+D_{eg}\left(\tau\right)\ket{e}\bra{g}+D_{ge}\left(\tau\right)\ket{g}\bra{e}+D_{gg}\left(\tau\right)\ket{g}\bra{g}\right)\right]}{\text{Tr}\left[\ket{\psi_{fI}}\bra{\psi_{fI}}e^{\mathcal{D}\tau}\left(\ket{\psi_i}\bra{\psi_i}\right)\right]},
\end{eqnarray} 
where
\begin{equation}
\hat{D}\left(\tau\right)=e^{\mathcal{D}\tau}\left(\hat{A}_{SI}\ket{\psi_i}\bra{\psi_i}\right)
\end{equation}
and the $D_{ij}\left(\tau\right)$ coefficients are the matrix elements of the operator $\hat{D}\left(\tau\right)$. The dissipator $\mathcal{D}$ represents the Lindbladian in the interaction picture. The observable $\hat{A}_{SI}$ is in principle evaluated in the interaction picture at time $t/2$. However, for the purpose of this discussion, we could neglect the small effect of the free evolution and consider that $\hat{A}_{SI}\left(t/2\right)=\hat{A}_{S}$ without loss of generality, since $t$ is a small constant parameter. The trace of $\hat{D}\left(\tau\right)$, the result of the evolution of the operator $\hat{A}_{SI}\ket{\psi_i}\bra{\psi_i}$, is preserved at all times $\tau$, i.e., $\text{Tr}\left(e^{\mathcal{D}\tau}\left(\hat{A}_{SI}\ket{\psi_i}\bra{\psi_i}\right)\right)=\text{Tr}\left(\hat{A}_{SI}\ket{\psi_i}\bra{\psi_i}\right)$. Given that the Lindbladian dynamics associated with dissipation drives any operator to the ground state in the limit of infinite time (in other words, $\lim_{\tau\to\infty}e^{\mathcal{D}\tau}(\hat{O})=\text{Tr}[\hat{O}]\ket{g}\bra{g}$), the coefficient that multiplies the ground density operator at long times should be equal to the expectation value of the operator $\hat{A}_{SI}$, i.e., $\lim_{\tau\to\infty}D_{gg}\left(\tau\right)=\text{Tr}\left(\hat{A}_{SI}\ket{\psi_i}\bra{\psi_i}\right)$. Therefore, the limit of the weak value can be expressed as
\begin{equation}
\label{eq:limit_weak_value}
\lim_{\tau \to\infty}A_{S,w}\left(\tau\right)=\frac{\text{Tr}\left[\ket{\psi_{fI}}\bra{\psi_{fI}}\text{Tr}\left(\hat{A}_{SI}\ket{\psi_i}\bra{\psi_i}\right)\ket{g}\bra{g}\right]}{|\braket{\psi_{fI}}{g}|^2}=\text{Tr}\left(\hat{A}_{SI}\ket{\psi_i}\bra{\psi_i}\right)=\bra{\psi_i}\hat{A}_{SI}\ket{\psi_i}.
\end{equation}

In conclusion, under dissipation, any weak value of a general observable approaches its expectation value on the initial state in the limit of infinite time. In particular, this result remains true regardless of the dimensionality of the quantum state and the chosen post-selected state, under the assumption of a unique steady state.

\subsection{Degenerate ground state}
We demonstrate that when a system's ground state is degenerate, amplification can occur at infinite dissipation times. Specifically, let us consider an $N$-level system that has a two-dimensional degenerate ground state. We chose the two orthogonal states $\ket{g_1}$ and $\ket{g_2}$ as a basis to describe the ground state. Depending on the initial state, the final evolution of the system can result in the final state being either $\ket{g_1}$, $\ket{g_2}$, a linear combination of the two, or a density matrix involving both states. In cases where the evolutions under dissipation in the denominator and numerator of the weak value yield the same final state, the weak value at infinite dissipation times tends towards the expectation value, as in Eq. (\ref{eq:limit_weak_value}). However, in cases where the final operators differ, amplification can occur at very long times. In this scenario, assuming pure pre- and post-selected states for simplicity, the weak value at infinite dissipation times can be expressed as follows,
\begin{eqnarray}
&& \mkern-36mu \lim_{\tau\rightarrow\infty}A_{S,w}\left(\tau\right) = \lim_{\tau\rightarrow\infty}\frac{\text{Tr}\left[\ket{\psi_{fI}}\bra{\psi_{fI}}e^{\mathcal{D}\tau}\left(\hat{A}_{SI}\ket{\psi_i}\bra{\psi_i}\right)\right]}{\text{Tr}\left[\ket{\psi_{fI}}\bra{\psi_{fI}}e^{\mathcal{D}\tau}\left(\ket{\psi_i}\bra{\psi_i}\right)\right]}\\ \nonumber
&&=\frac{a_{11}\big|\braket{\psi_{fI}}{g_1}\big|^2+a_{22}\big|\braket{\psi_{fI}}{g_2}\big|^2+a_{21}\braket{\psi_{fI}}{g_2}\braket{g_1}{\psi_{fI}}+a_{12}\braket{\psi_{fI}}{g_1}\braket{g_2}{\psi_{fI}}}{b_{11}\big|\braket{\psi_{fI}}{g_1}\big|^2+b_{22}\big|\braket{\psi_{fI}}{g_2}\big|^2+b_{21}\braket{\psi_{fI}}{g_2}\braket{g_1}{\psi_{fI}}+b_{12}\braket{\psi_{fI}}{g_1}\braket{g_2}{\psi_{fI}}},
\end{eqnarray}
where
\begin{eqnarray}
&&\lim_{\tau\rightarrow\infty}e^{\mathcal{D}\tau}\left(\hat{A}_{SI}\ket{\psi_i}\bra{\psi_i}\right)=\sum_{j=1}^2\sum_{k=1}^2 a_{jk}\ket{g_j}\bra{g_k}, \\ 
&&\lim_{\tau\rightarrow\infty}e^{\mathcal{D}\tau}\left(\ket{\psi_i}\bra{\psi_i}\right)=\sum_{j=1}^2\sum_{k=1}^2 b_{jk}\ket{g_j}\bra{g_k}.
\end{eqnarray}
One can find amplification and complex weak values even at infinite time by choosing the appropriate post-selected state and observable. In that case, since these quantities cannot be any expectation value of the operator $\hat{A_S}$, the weak value is considered anomalous.

An alternative explanation stems from the decomposition of the unnormalized state $\hat{A}_{SI}\ket{\psi_i}=\left\langle\hat{A}_{SI}\right\rangle_i\,\ket{\psi_i} + \Delta_i \hat{A}_{SI}\, \ket{\psi_i^\perp}$, where the average is $\left\langle\hat{A}_{SI}\right\rangle_i=\bra{\psi_i}\hat{A}_{SI}\ket{\psi_i}$, the quantum uncertainty is $\Delta_i \hat{A}_{SI}=\sqrt{\left\langle\hat{A}_{SI}^2\right\rangle_i-\left\langle\hat{A}_{SI}\right\rangle_i^2}$, and the normalized state $\ket{\psi_i^\perp}$ is orthogonal to the initial state $\ket{\psi_i}$. Therefore, considering the linearity of the dissipator evolution, we can write the weak value as
\begin{equation}
    A_{S,w}\left(\tau\right) = \left\langle\hat{A}_{SI}\right\rangle_i + \Delta_i \hat{A}_{SI} \ \frac{\text{Tr}\left[\ket{\psi_{fI}}\bra{\psi_{fI}}e^{\mathcal{D}\tau}\left(\ket{\psi_i^\perp}\bra{\psi_i}\right)\right]}{\text{Tr}\left[\ket{\psi_{fI}}\bra{\psi_{fI}}e^{\mathcal{D}\tau}\left(\ket{\psi_i}\bra{\psi_i}\right)\right]},
\end{equation}
where the second term is responsible for any anomalousness of the weak value. We note that the trace of the coherence $\ket{\psi_i^\perp}\bra{\psi_i}$ is zero. When the ground state is not degenerate, the operator $e^{\mathcal{D}\tau}\left(\ket{\psi_i^\perp}\bra{\psi_i}\right)$ can only decay to $0$ in the ground state and this second term does not contribute to the weak value. As explained previously, the weak values then decays to the average value. However, when the ground state is degenerate, depending on the nature of the various dissipation channels, the non-Hermitian operator $\ket{\psi_i^\perp}\bra{\psi_i}$ can decay to a mixture of the zero operator and coherences in the ground state manifold. The part that decays to zero does not contribute to the weak value in the limit of infinite dissipation time, similarly to the non-degenerate case. However, the part that decays to non-Hermitian coherences of the type $\ketbra{g_m}{g_n}$ (with $\langle g_m\vert g_n\rangle=0$) has trace zero but does contribute to the weak value, even in the limit of infinite dissipation time. This argument can easily be generalized to mixed initial states, beyond the pure state case examined here. Analyzing the evolution of the coherence  $e^{\mathcal{D}\tau}\left(\ket{\psi_i^\perp}\bra{\psi_i}\right)$ under dissipation may help find specific initial and final states to evidence anomalous weak values in degenerate systems.

\begin{figure}[t]
 \centering
\includegraphics[width=0.65\textwidth]{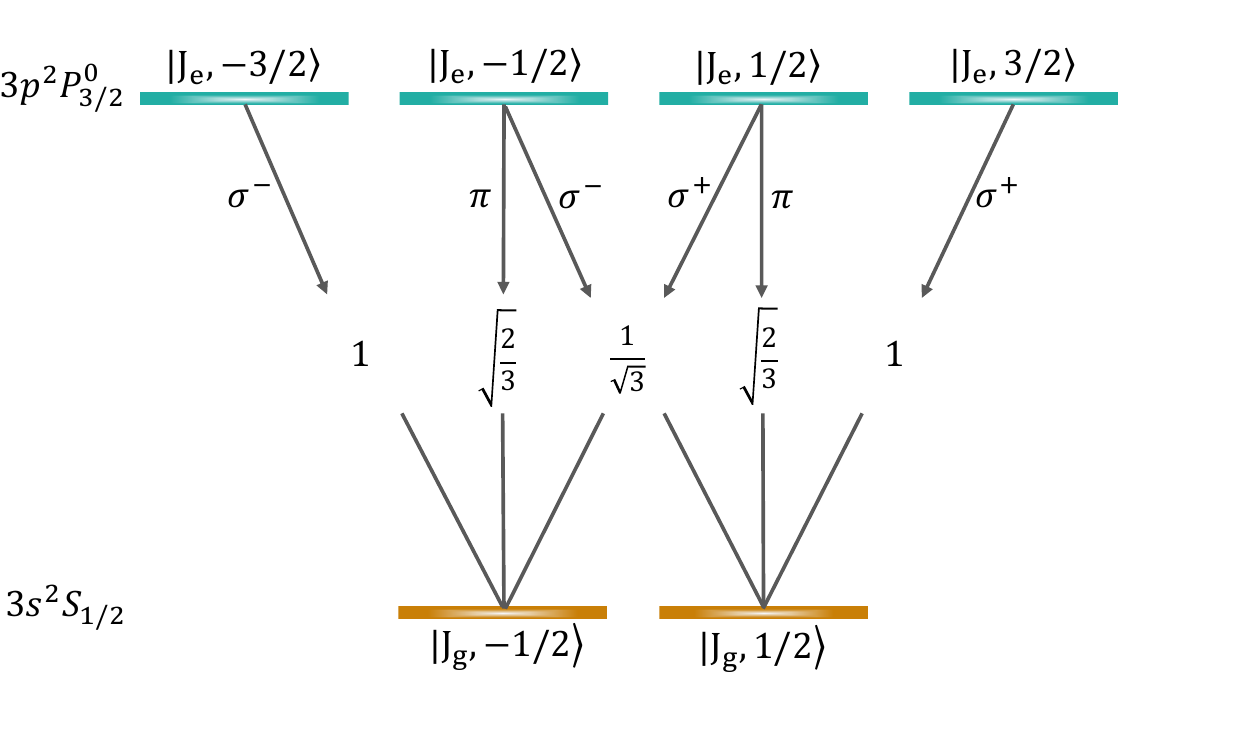}
\caption{Atomic transition $J_g=\frac{1}{2} \leftrightarrow J_e=\frac{3}{2}$. The Clebsch-Gordan coefficients for each transition, $C^{J_g,1,J_e}_{m_g,q,m_e}=\braket{J_g,m_g;1,q}{J_e,m_e}$ are shown for each allowed transition.\label{fig:scheme_sodium}}
\end{figure}

As an example to demonstrate the anomalous properties of weak values in systems with degenerate ground states, we examine a simple case involving a sodium atom. Specifically, we consider a situation in which the orbitals $1s$, $2s$, and $2p$ are all fully occupied, and there is one electron in the degenerate level $3s$. When this electron is excited, it can undergo a transition to the nearby $3p$ level. In particular, one of the most intense transitions is the $J_g=\frac{1}{2} \leftrightarrow J_e=\frac{3}{2}$ transition that produces the main spectral line in the sodium doublet. In this system, the ground state is degenerate, while there are four possible excited states, making it a six-level system, Fig.~\ref{fig:scheme_sodium}.

The Lindbladian governing the de-excitation of the atom can be expressed in the interaction picture as 
\begin{equation}
\label{eq:L_I}
\mathcal{D}\left(\hat{\sigma}_I\right)=\Gamma\sum_{q=0,\pm}\left(\hat{L}_q\hat{\sigma}_I\hat{L}^{\dagger}_q-\frac{1}{2}\{\hat{L}_q^{\dagger}\hat{L}_q,\hat{\sigma}_I\}\right),
\end{equation}
where $\Gamma$ is a characteristic spontaneous emission rate for the transition $J_g=\frac{1}{2}\leftrightarrow J_e=\frac{3}{2}$, and
\begin{eqnarray}
\hat{L}_q\ket{J_e, m_e=m_g+q}=C^{J_g,1,J_e}_{m_g,q,m_e}\ket{J_g,m_g}, \hspace{1 cm} \hat{L}_q\ket{J_g,m_g}=0.
\end{eqnarray}
For more details on the expression of $\hat{L}_q$, see appendix \ref{appendix:expression_l_q}. To give a concrete example of the anomalous weak value generated by dissipation, let us consider the following pre- and post-selected states, 
\begin{eqnarray}
\ket{\psi_i}&=&\frac{1}{2}\left(\ket{J_e,-3/2}+i\ket{J_e,-1/2}+\ket{J_e,1/2}+\ket{J_e,3/2}\right),\\ \nonumber
\ket{\psi_{fI}}&=&\alpha\ket{J_e,-3/2}+-0.995\ket{J_e,-1/2}-\alpha\left(1+i\right)\ket{J_e,3/2}+\alpha\ket{J_g,-1/2}\\ \nonumber
&& + \left(-0.00734+0.00114i\right)\ket{J_g,1/2}, 
\end{eqnarray}
with $\alpha=0.0498$. We choose again a state that is constant in the interaction picture to analyze solely the effects of the dissipation. The chosen observable to be measured is the angular momentum $\hat{A}_{S}\approx\hat{A}_{SI}\left(t/2\right)=\hat{J}_y$, or by setting $\hbar=1$,
\begin{equation}
\hat{J}_y=
\begin{pmatrix}
0 & i\frac{\sqrt{3}}{2} & 0 & 0 & 0 & 0\\
-i\frac{\sqrt{3}}{2} & 0 & i & 0 & 0 & 0 \\
0 & -i & 0 & i\frac{\sqrt{3}}{2} & 0 & 0 \\
0 & 0 & -i\frac{\sqrt{3}}{2} & 0 & 0 & 0\\
0 & 0 & 0 & 0 & 0 & \frac{i}{2}\\
0 & 0 & 0 & 0 & -\frac{i}{2} & 0
\end{pmatrix},
\end{equation}
in a basis ordered by decreasing magnetic quantum number $m$ starting from the four excited states and ending with the two ground states (see Fig.~\ref{fig:scheme_sodium}).

\begin{figure}[t]
\centering
\includegraphics[width=0.7\textwidth]{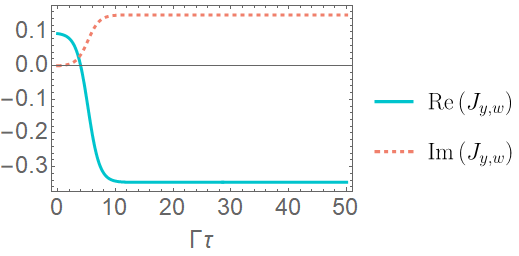}
\caption{Real and imaginary parts of the weak value of $\hat{J}_y$ as a function of the product of the dissipation time and the dissipation rate. The pre- and post-selected states are detailed in the text and chosen to illustrate the occurrence of an increasing anomaly due to dissipation.}
\label{fig:representation_weak_value_time_degenerate_1}
\end{figure}

Using the chosen pre- and post-selected states and observable, we find that the weak value without dissipation is $A_w(\tau=0)=0.0954$. This weak value is not anomalous, as its imaginary part is zero and its modulus, $0.0954$, lies in the range of the spectrum of $\hat{J}_y$, whose smallest and largest eigenvalues are $\pm\frac{3}{2}$. At infinite dissipation time, the modulus and the imaginary part of the weak value increase in magnitude to $A_w(\tau\rightarrow\infty)=-0.346 + 0.151i$. The dissipation generates an anomalous behavior of the weak value, by increasing the imaginary part from $0$ to $0.151$. In Fig.~\ref{fig:representation_weak_value_time_degenerate_1}, we show the evolution of the real and imaginary parts of the weak value as a function of the product $\Gamma \tau$ of dissipation time and dissipation rate. One can appreciate that, at very long times, the imaginary part of the weak value is not zero and the real part is larger in modulus than it was at null dissipation time.

\begin{figure}[t]
\centering
\includegraphics[width=0.7\textwidth]{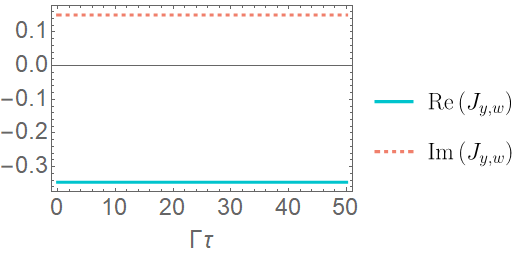}
\caption{Real and imaginary parts of the weak value of $\hat{J}_y$ as a function of the product of the dissipation time and the dissipation rate. The pre- and post-selected states are detailed in the text and are chosen to create a constant anomalous weak value.}\label{fig:representation_weak_value_time_degenerate_2} 
\end{figure}

We can also obtain an anomalous weak value that is preserved and completely constant over time, by choosing the following pre- and post-selected states for the same system:
\begin{eqnarray}
\label{eq:second_pair_pre_and_post_Selected_degenerate}
\ket{\psi_i}&=&\frac{1}{2}\left(\ket{J_e,-3/2}+i\ket{J_e,-1/2}+\ket{J_e,1/2}+\ket{J_e,3/2}\right),\\ \nonumber
\ket{\psi_{fI}}&=&0.989\ket{J_g,-1/2}+\left(-0.146+0.0226i\right)\ket{J_g,1/2}.
\end{eqnarray}
Despite the orthogonality of the pre- and post-selected states, as defined in Eq.~(\ref{eq:second_pair_pre_and_post_Selected_degenerate}), the presence of dissipation in the system ensures that the weak value denominator is non-zero for $\tau>0$.
Figure~\ref{fig:representation_weak_value_time_degenerate_2} depicts the evolution of the real and imaginary part of the weak value of $\hat{J}_y$. As one can see, the weak value is constant and has an imaginary part different from $0$.

Anomalous weak values, which are quantities different from any possible expectation value (i.e., larger than the maximum expectation value, smaller than the minimal expectation value, or complex values), are linked to contextuality \cite{pusey2014anomalous, kunjwal2019anomalous}, a non-classical property. Dissipation generally destroys the quantum superposition and coherence of a system, and without these quantum properties, there is no amplification through anomalous weak values. However, in systems with a degenerate ground state, the final state can still present quantum superposition and coherence, which allows us to maintain the anomalous character of the weak value even in the limit of infinite dissipation times. In non-degenerate systems, dissipation inevitably eliminates the anomalous properties of weak values over time, preventing amplification in this regime. Nonetheless, weak values can still be leveraged to extract information about the system evolution over short dissipation times. During the early stages of dissipation, the Lindbladian can be approximated by a Taylor series, enabling us to extract parameters related to the dynamics from the weak value evolution. In the upcoming sections, we will examine a few examples.

\section{Weak measurement in the Rabi model}\label{sec:RabiModel}
\begin{figure}[tbh]
 \centering
\includegraphics[width=1.0\textwidth]{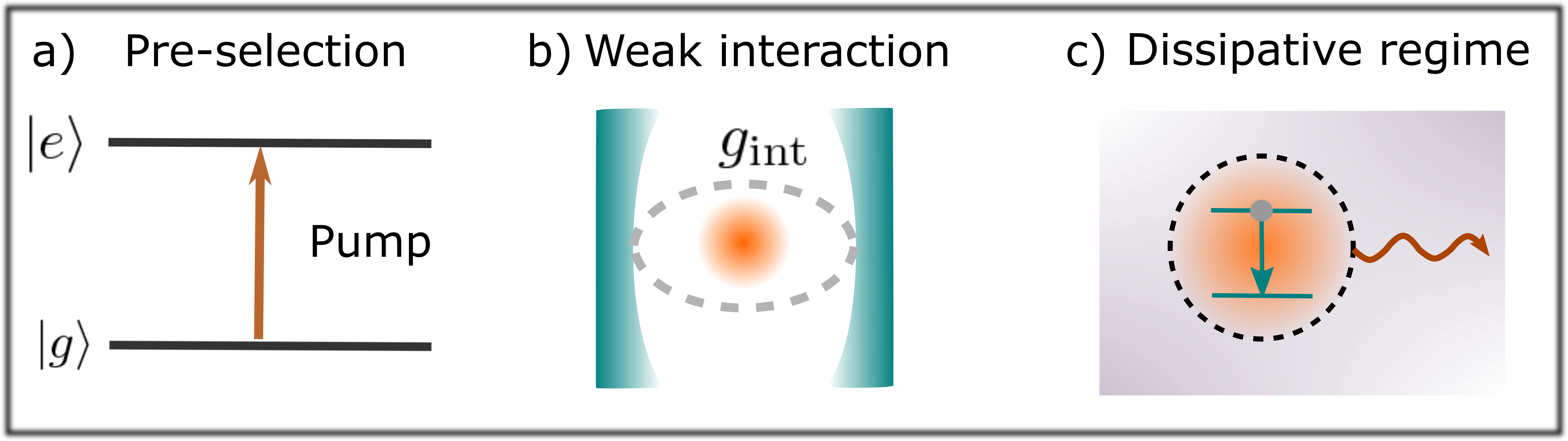}
\caption{The weak measurement under dissipation scheme involves four stages. Firstly, pre-selection of the system is achieved by pumping the atom, as shown in (a), leaving it in a chosen superposition of the ground and excited states. Secondly, a weak interaction occurs in a closed single-mode cavity, as depicted in (b). Thirdly, after the weak interaction, the atom undergoes dissipation, as shown in (c). Finally, post-selection is performed on the atom. \label{fig:Rabi_weak_measurement}}
\end{figure}

In this section, we apply the theoretical principles of weak measurements under dissipation to a specific setup involving a two-level atom system. In this system, a weak measurement of the internal state of the atom is performed using the field of a cavity. The cavity mode serves as the meter or ancilla, and the atom-cavity dynamics are governed by the Rabi model. Figure~\ref{fig:Rabi_weak_measurement} illustrates the various steps in this process.  Initially, the atom is pre-selected, for example by pumping it with a laser pulse of appropriate frequency and intensity, leaving it in a superposition of its ground and excited states, as shown in (a). Following pre-selection, the weak interaction occurs with a cavity mode whose initial quantum state is known, as shown in (b). Dissipation then occurs as soon as the atom leaves the cavity, as a result of its interaction with a quantized radiation field (either the free field or the field of a second lossy cavity into which it moves) acting as an environment with an infinite number of degrees of freedom, as shown in (c). Finally, post-selection is performed on the atom and the bosonic cavity mode is read out. By measuring the field quadratures of the cavity mode, $\hat{Q}$ and $\hat{P}$, we can extract the real and imaginary components of the weak value from the shift in the expectation value of the quadratures. 

When the atom is in the cavity, its interaction with the field mode is described by the Rabi model, corresponding to the atom-field Hamiltonian $\hat{H}=\hat{H}_{\text{atom}}+\hat{H}_{\text{field}}+\hat{H}_{\mathrm{int}}$ with
\begin{eqnarray}
\label{eq:hamiltonians_rabi_model}
\hat{H}_{\text{atom}}&=&\frac{1}{2}\hbar\omega_a\hat{\sigma}_z, \\ \nonumber
\hat{H}_{\text{field}}&=&\hbar\omega_f \hat{a}^{\dagger}\hat{a},\\ \nonumber
\hat{H}_{\mathrm{int}}&=&\hbar g_{\mathrm{int}}\hat{\sigma}_x\otimes\left(\hat{a}^\dagger+\hat{a}\right).
\end{eqnarray} 
where $\omega_a$ and $\omega_f$ are respectively the frequencies of the atom and the cavity, $g_{\mathrm{int}}$ is the atom-cavity coupling constant, $\hat{a}^{\dagger}$ and $\hat{a}$ are the creation and annihilation operators of the field, and $\hat{\sigma}_j$ the Pauli matrices. 
The field quadratures are defined by $\hat{Q}=\sqrt{\frac{\hbar}{2\omega_f}}\left(\hat{a}^{\dagger}+\hat{a}\right)$ and $\hat{P}=i\sqrt{\frac{\hbar\omega_f}{2}}\left(\hat{a}^{\dagger}-\hat{a}\right)$. In this model, the pointer $\hat{N}$ from Eq.~(\ref{eq:interaction_hamiltonian}) is thus given by $\hat{N}=\sqrt{\frac{2 \omega_f}{\hbar}}\hat{Q}$. ($\hat{N}$ should not to be confused here with the number operator $\hat{a}^\dagger\hat{a}$.)

\subsection{Usual weak measurement approximation}

To begin, the atomic system is pre-selected in the state $\hat{\sigma}_i$. The meter initial state is denoted by $\hat\mu_0$. For an arbitrary initial state of the meter, the weak measurement results are provided by the general expressions (\ref{eq:expectation_value_L_dissipation_WithDenom}) or (\ref{eq:expectation_value_L_RealCommutator-ImaginaryAntiCommutator_WithDenominatorAdded}). In order to use the corresponding simpler results (\ref{eq:expectation_value_L_dissipation}) or (\ref{eq:expectation_value_L_RealCommutator-ImaginaryAntiCommutator}) for the measurement average, we can require that $\langle \hat{N}_I\left(t/2\right)\rangle_0=0$; we can also take advantage of (\ref{eq:expectation_value_L_dissipation_without_commutator}) when the initial meter state commutes with $\hat{H}_{\mathrm{field}}$. The simplest meter state to consider is the vacuum state, denoted by $\ket{0}\bra{0}$, but the meter shifts observed with energy eigenstates, coherent states, squeezed states, and thermal-equilibrium states can all be evaluated straightforwardly, enabling the determination of the weak value. In the following, we will determine the meter shifts for a general meter state at first. Then, we will discuss a few simple examples.

The observable weakly measured as a result of the interaction $\hat{H}_{\mathrm{int}}$ (\ref{eq:hamiltonians_rabi_model}) of the Rabi model is the operator $\hat{\sigma}_x$, at least if we assume an instantaneous weak interaction (equivalent to adding a Dirac distribution $\delta\left(t\right)$ in the interaction Hamiltonian). For a short --but not instantaneous-- interaction time with respect to the free evolution of the atom ($\omega_a t \ll 1$) and cavity field ($\omega_f t \ll 1$), we can use the weak measurement approximation (\ref{eq:interactionHamiltonianApprox})--(\ref{eq:evolution_rho_interaction_picture_after_short_time_simplified}) on which relies our general theory. In that case, we showed that the operator effectively probed is 
\begin{equation}\label{eq:mesuredOperator-RabiModel}
    \hat{\sigma}_{xI}\left(t/2\right)=e^{\frac{i}{4} \omega_a t \hat\sigma_z}\hat{\sigma}_x e^{- \frac{i}{4} \omega_a t \hat\sigma_z}= \cos \left(\frac{\omega_a t}{2}\right)\hat{\sigma}_x - \sin \left(\frac{\omega_a t}{2}\right) \hat{\sigma}_y.
\end{equation}
We see that it corresponds to a small clockwise rotation of the $\hat{\sigma}_{x}$ operator around the $z$ axis. We will use the notation $\hat{\sigma}_{xI} = \vec{n}_I\cdot\hat{\vec{\sigma}}$ to represent this operator, with $\vec{n}_I=(\cos \frac{\omega_a t}{2},-\sin \frac{\omega_a t}{2},0)$ a three-dimensional real vector representing $\hat{\sigma}_{xI}$ on the Bloch sphere. 

We use Eq.~(\ref{eq:expectation_value_L_RealCommutator-ImaginaryAntiCommutator_WithDenominatorAdded}) to determine the shifts in terms of the real and imaginary parts of the weak value. Therefore, we need to compute the meter observables $\hat{N}_I$ and $\hat{L}_I$ in the interaction picture, as well as the commutator $\left[\hat{L}_I,\hat{N}_I\right]$ and anti-commutator $\left\{\hat{L}_I,\hat{N}_I\right\}$ for both cases $\hat{L}_I=\hat{Q}_I$ and $\hat{L}_I=\hat{P}_I$, since the two field quadratures serve as our conjugate meter observables. In the interaction picture, the creation and annihilation operators are represented as $\hat{a}_I\left(t^\prime\right)=e^{- i \omega_f t^\prime}\hat{a}$ and $\hat{a}_I^\dagger\left(t^\prime\right)=e^{i \omega_f t^\prime}\hat{a}^\dagger$. Thus, we can promptly determine the observables
\begin{eqnarray}\label{eq:Rabi-NI}
  \hat{N}_I\left(t/2\right)&=&\left[\hat{a}^\dagger e^{i \omega_f \left(t/2\right)}+\hat{a} e^{-i \omega_f \left(t/2\right)}\right],\\\label{eq:Rabi-QI}
    \hat{Q}_I\left(t+\tau\right)&=&\sqrt{\frac{\hbar}{2 \omega_f}}\left[\hat{a}^\dagger e^{i \omega_f \left(t+\tau\right)}+\hat{a} e^{-i \omega_f \left(t+\tau\right)}\right], \\\label{eq:Rabi-PI}
    \hat{P}_I\left(t+\tau\right)&=&\sqrt{\frac{\hbar \omega_f}{2}} i \left[\hat{a}^\dagger e^{i \omega_f \left(t+\tau\right)}-\hat{a} e^{-i \omega_f \left(t+\tau\right)}\right],
\end{eqnarray}
the commutators 
\begin{eqnarray}\label{eq:Rabi-CommutatorQN}
  \left[\hat{Q_I}\left(t+\tau\right),\hat{N}_I\left(t/2\right)\right]&=&-2 i \sqrt{\frac{\hbar}{2 \omega_f}} \sin\left[ \omega_f\left(t/2+\tau\right)\right]\ \hat{\mathbb{1}}, \\\label{eq:Rabi-CommutatorPN}
    \left[\hat{P_I}\left(t+\tau\right),\hat{N}_I\left(t/2\right)\right]&=&-2 i \sqrt{\frac{\hbar \omega_f}{2}} \cos\left[  \omega_f\left(t/2+\tau\right)\right]\ \hat{\mathbb{1}},
\end{eqnarray}
as well as the anti-commutators
\begin{eqnarray}\label{eq:Rabi-AntiCommutatorQN}
  \left\{\hat{Q_I}\left(t+\tau\right),\hat{N}_I\left(t/2\right)\right\}&=&2 \sqrt{\frac{\hbar}{2 \omega_f}} \cos\left[  \omega_f\left(t/2+\tau\right)\right]\ \left(2 \hat{a}^\dagger\hat{a}+\hat{\mathbb{1}}\right) \nonumber\\
  & + & 2 \sqrt{\frac{\hbar}{2 \omega_f}} \left[\hat{a}^{\dagger 2} e^{i \omega_f \left(3t/2+\tau\right)}+\hat{a}^2 e^{-i \omega_f \left(3t/2+\tau\right)}\right], \\\label{eq:Rabi-AntiCommutatorPN}
    \left\{\hat{P_I}\left(t+\tau\right),\hat{N}_I\left(t/2\right)\right\}&=&- 2  \sqrt{\frac{\hbar \omega_f}{2}} \sin\left[   \omega_f\left(t/2+\tau\right)\right]\ \left(2 \hat{a}^\dagger\hat{a}+\hat{\mathbb{1}}\right)\nonumber\\
     & + & 2 i\sqrt{\frac{\hbar \omega_f}{2}} \left[\hat{a}^{\dagger 2} e^{i \omega_f \left(3t/2+\tau\right)}-\hat{a}^2 e^{-i \omega_f \left(3t/2+\tau\right)}\right].
\end{eqnarray}
Now, we can simply evaluate the expectation values of these quantities in the meter initial state to determine the meter shifts (\ref{eq:expectation_value_L_RealCommutator-ImaginaryAntiCommutator_WithDenominatorAdded}). Interestingly, we can already observe that the expectation value of the commutators (\ref{eq:Rabi-CommutatorQN}) and (\ref{eq:Rabi-CommutatorPN}) do not depend on the meter initial state since they are proportional to the identity. This means that the initial meter state does not influence the meter shift arising from the real part of the weak value (at least when we can neglect the effect of the denominator in the shift expression). In contrast, the expectation value of the anti-commutators depends on the initial meter state. Thus, the choice of the initial meter state will influence the meter shift associated to the imaginary part of the weak value. It is worth noticing that the expectation value of the operator $2 \hat{a}^\dagger\hat{a}+\hat{\mathbb{1}}$ present in (\ref{eq:Rabi-AntiCommutatorQN}) and (\ref{eq:Rabi-AntiCommutatorPN}) is proportional to the average energy in the initial meter state (including the zero-point energy term). Using an initial meter state with an average energy larger than the energy of the vacuum state will thus generally increase the meter shift resulting from the imaginary part of the weak value, relatively to the shift due to the real part (at least, when we can ignore the contributions of the $\hat{a}^2$ and $\hat{a}^{\dagger 2}$ terms). 

For simplicity, let us now assume that the meter initial state is an energy eigenstate $\hat{\mu}_0=\ketbra{n}{n}$. In that case, all the expectation values of the meter observables $\hat{N}_I$, $\hat{Q}_I$, and $\hat{P}_I$ (\ref{eq:Rabi-NI}--\ref{eq:Rabi-PI}) are 0, while terms in $\hat{a}^2$ and $\hat{a}^{\dagger 2}$ in (\ref{eq:Rabi-AntiCommutatorQN}) and (\ref{eq:Rabi-AntiCommutatorPN}) average to zero as well. As a result, we find the following meter shifts at the end of the weak measurement:
\begin{eqnarray}\label{eq:Rabi-Qfn}
    \langle\hat{Q}\rangle_f =-2 g t \sqrt{\frac{\hbar}{2 \omega_f}} \left[\sin{\omega_f \left( t/2+\tau\right)}\ \textrm{Re} \sigma_{S,w}\left(\tau\right)-\left(2 n +1\right)\cos{\omega_f \left( t/2+\tau\right)}\ \textrm{Im} \sigma_{S,w}\left(\tau\right)\right], \\\label{eq:Rabi-Pfn}
     \langle\hat{P}\rangle_f =-2 g t \sqrt{\frac{\hbar \omega_f}{2 }} \left[\cos{\omega_f \left( t/2+\tau\right)}\ \textrm{Re} \sigma_{S,w}\left(\tau\right)+\left(2 n +1\right)\sin{\omega_f \left( t/2+\tau\right)}\ \textrm{Im} \sigma_{S,w}\left(\tau\right)\right].
\end{eqnarray}
As just discussed before, the shifts proportional to the real part of the weak value are identical for all the energy eigenstates, while the shifts proportional to the imaginary part increase linearly with the energy level $n$. In these expressions, the weak value $\sigma_{S,w}\left(\tau\right)$ is the weak value of the observable $\vec{n}_I\cdot\hat{\vec{\sigma}}$ (\ref{eq:mesuredOperator-RabiModel}) under dissipation, defined as in Eq.~(\ref{eq:weak_value_with_dissipation}) by imposing $\hat{A}_{SI}=\hat{\sigma}_{xI}$. The expressions (\ref{eq:Rabi-Qfn}) and (\ref{eq:Rabi-Pfn}) would remain essentially identical for a meter initial state in thermal equilibrium: in that case, the energy level $n$ should be replaced by the average energy level in the thermal state $n_{\mathrm{eq}}(T)$, with $2 n_{\mathrm{eq}}+1=\coth{\left[\hbar \omega_f/\left(2 k_B T\right)\right]}$ where $k_B$ is the Boltzmann constant and $T$ the temperature. We see thus that the meter shift due to the imaginary part of the weak value is temperature dependent, while the shift due to the real part is not. For a squeezed vacuum state as the meter initial state, it is theoretically possible to amplify or attenuate exponentially the shift due to the imaginary part of the weak value as a function of the squeezing parameter $z$ (with a factor $e^{\pm 2 \left\vert z \right\vert}$ replacing $(2n+1)$ in (\ref{eq:Rabi-Qfn}) and (\ref{eq:Rabi-Pfn}) for the particular argument $\arg z = \omega_f t$; also, because the terms $\hat{a}^2$ and $\hat{a}^{\dagger 2}$ contribute to the anti-commutator average for squeezed states, the factor $e^{\pm 2 \left\vert z \right\vert}$ is not proportional to the mode average energy $\sinh{\left\vert z \right\vert}$). When choosing a coherent state for the meter initial state, the shift expressions remain straightforward to compute. However, they become much less practical to handle because the expectation values of the meter observables $\hat{N}_I$, $\hat{Q}_I$, and $\hat{P}_I$ are generally non-zero and then contribute to (\ref{eq:expectation_value_L_RealCommutator-ImaginaryAntiCommutator_WithDenominatorAdded}) by adding a term independent on the weak value on the numerator and by preserving the term proportional to the imaginary part of the weak value in the denominator. From that perspective, coherent states do not seem ideal to retrieve the weak values from the experimental meter shifts.
 
Now, we turn our attention to the dependence of the meter shifts on the dissipation duration $\tau$, which determines the post-selection time. For reasons that will soon become apparent, we define the polar representation of the weak value \cite{mc2016, mc2017, lbf2022} by $\sigma_{S,w}\left(\tau\right)=\left\vert \sigma_{S,w}\left(\tau\right) \right\vert e^{i \varphi_w(\tau)}$, where $\left\vert \sigma_{S,w}\left(\tau\right) \right\vert$ is its modulus and $\varphi_w(\tau)$ is its argument as a function of the dissipation duration $\tau$. We consider that the meter initial state is the vacuum state for simplicity. After setting $n=0$ in (\ref{eq:Rabi-Qfn}) and (\ref{eq:Rabi-Pfn}), the expectation values of the $\hat{Q}$ and $\hat{P}$ field quadratures become
\begin{eqnarray}\label{eq:QRabiQuadrature}
\langle\hat{Q}\rangle_{f}&=&2 g_{\mathrm{int}} t \sqrt{\frac{\hbar}{2\omega_f}} \left\vert \sigma_{S,w}\left(\tau\right) \right\vert \sin{\left[\varphi_w(\tau)-\omega_f (t/2+\tau)\right]}, \\ \label{eq:PRabiQuadrature}
\langle\hat{P}\rangle_{f}&=&-2 g_{\mathrm{int}} t \sqrt{\frac{\hbar \omega_f}{2}} \left\vert \sigma_{S,w}\left(\tau\right) \right\vert \cos{\left[\varphi_w(\tau)-\omega_f (t/2+\tau)\right]}.
\end{eqnarray}
 In the limit of negligible free evolution of the meter, i.e.\ when $\omega_f (t/2+\tau)\approx 0$, the expectation value of the $\hat{Q}$ field quadrature is proportional to the imaginary part of the weak value $\left\vert \sigma_{S,w}\left(\tau\right) \right\vert \sin \varphi_w(\tau)$, as seen from Eq.~(\ref{eq:expectation_value_N}). In the
 this same limit, the expectation value of the $\hat{P}$ quadrature, associated with the canonical conjugate of the weak measurement pointer, is proportional to the real part of the weak value $\left\vert \sigma_{S,w}\left(\tau\right) \right\vert \cos \varphi_w(\tau)$, as seen from Eq.~(\ref{eq:expectation_value_M}). 
When the meter evolution due to its Hamiltonian $\hat{H}_{\text{field}}$ can be neglected, the shifts in position and momentum are proportional to the real and imaginary parts of the weak value, respectively. When the contributions from the term $\omega_f (t/2+\tau)$ cannot be neglected in Eqs.~(\ref{eq:QRabiQuadrature}) and (\ref{eq:PRabiQuadrature}), then the weak value rotates with time $\tau$ in the meter phase space defined by the $\hat{Q}$--$\hat{P}$ quadratures (assuming that the contribution $\omega_f \tau$ varies much faster than the weak value argument $\varphi_w(\tau)$). This apparent rotation of the weak value results purely from the free meter evolution with time and does not arise from a modification of the weak value itself. Consequently, practical measurements need to be implemented differently for cases where the weak value evolution is much faster than the free-meter evolution, and for cases where the free-meter evolution significantly outpaces the weak value evolution. Nevertheless, in both situations, by knowing the parameters of the weak measurement, $t$ and $g_{\mathrm{int}}$, we can determine the weak value as a function of the dissipation time $\tau$ from the weak measurement results and, in turn, use it to extract information about the dissipative evolution. Even for more complex initial meter states, it is possible to invert the expressions linking the two meter expectation values to the complex weak value (see Appendix \ref{appendix:WVfromWM}).

In the upcoming sections, we examine how to use the weak value expression to extract information regarding the dissipative dynamics of the atom. In this section, we provide a more comprehensive treatment of the weak measurement in the presence of dissipation, illustrating the general theory in depth.

\subsection{General expression of the weak value for two-level systems with dissipation}
The Rabi model allows us to investigate analytically how the weak value evolves with the dissipation duration $\tau$. Indeed, taking a single dissipation channel characterized by the time-independent damping constant $\gamma$ and the jump operator $\hat\sigma_{-}$ in the dissipator $\mathcal{D}$ defined in (\ref{eq:Lindblad_master_equation}), an arbitrary matrix $C=c_{gg}\ketbra{g}{g}+c_{ge}\ketbra{g}{e}+c_{eg}\ketbra{e}{g}+c_{ee}\ketbra{e}{e}$ becomes
\begin{equation}\label{eq:dissipationIn2lvlMatrix}
    e^{\mathcal{D}\tau}\left(C\right)=
    \begin{pmatrix}
          c_{ee} e^{-\gamma \tau} & c_{eg} e^{-\frac{1}{2}\gamma \tau}\\
        c_{ge} e^{-\frac{1}{2}\gamma \tau} & c_{gg}+c_{ee}\left(1-e^{-\gamma \tau}\right) \\
    \end{pmatrix}.
\end{equation}
Therefore, computing the weak value under dissipation (\ref{eq:weak_value_with_dissipation}) becomes straightforward. We choose arbitrary initial and final states characterized by the three-dimensional, real Bloch sphere vectors $\vec{i}$ and $\vec{f}$, respectively: $\hat\sigma_i=\frac{1}{2}(\hat{\mathbb{1}}+\vec{i}\cdot\hat{\vec{\sigma}})$ and $\hat\sigma_f=\frac{1}{2}(\hat{\mathbb{1}}+\vec{f}\cdot\hat{\vec{\sigma}})$. Unit vectors corresponds to arbitrary pure states of the atom, while vectors of length $<1$ describe arbitrary mixed states of the two-level system. In the weak value expression, the post-selected state should be evaluated in the interaction representation at time $t+\tau$. In particular, $\hat{\sigma}_{fI}\left(t+\tau\right)=\left\{\cos\left[{\frac{\omega_a}{2}\left(t+\tau\right)}\right] \ \hat{\mathbb{1}}+i\sin\left[{\frac{\omega_a}{2}\left(t+\tau\right)}\right]\ \hat{\sigma}_z\right\}\hat{\sigma}_f \left\{\cos\left[{\frac{\omega_a}{2}\left(t+\tau\right)}\right]\ \hat{\mathbb{1}}-i\sin\left[{\frac{\omega_a}{2}\left(t+\tau\right)}\right]\ \hat{\sigma}_z\right\}$. If we denote $\vec{f}=(f_x, f_y, f_z)$, then $\hat{\sigma}_{fI}$ is described by the vector
\begin{equation}\label{eq:RabiPostSelectedRotation}
    \vec{f}_I\left(t+\tau\right)=
    \begin{pmatrix}
        f_{Ix} \\
        f_{Iy} \\
        f_{Iz}
    \end{pmatrix}=
    \begin{pmatrix}
        f_x \cos \left[\omega_a \left(t+\tau \right)\right]+f_y \sin \left[\omega_a \left(t+\tau \right)\right]\\
        f_y \cos \left[\omega_a \left(t+\tau \right)\right]-f_x \sin \left[\omega_a \left(t+\tau \right)\right]\\
        f_z
    \end{pmatrix},
\end{equation}
which corresponds to a clockwise rotation of the initial vector $\vec{f}$ over time around the $z$ axis of the Bloch sphere. We thus see that the actual time of post-selection influences the state that is effectively post-selected in practice. Since the weakly measured operator is given by (\ref{eq:mesuredOperator-RabiModel}) and characterized by the vector $\vec{n}_I$, we now have the following expression for the weak value
\begin{equation}\label{eq:WVdissipation2lvlAnalytical}
    \hat{\sigma}_{S,w}\left(\tau\right)=\frac{
    \vec{f_I^\gamma}\cdot\vec{n}_I+\vec{i}\cdot\vec{n}_I \left(1+f_{Iz}^\gamma-f_{Iz}\right)+ i \vec{f_I^\gamma}\cdot \left(\vec{n}_I\cross\vec{i}\right)
    }{1+\vec{f_I^\gamma}\cdot\vec{i}+\left(f_{Iz}^\gamma-f_{Iz}\right)},
\end{equation}
where the vector $\vec{f_I^\gamma}$ is defined as
\begin{equation}\label{eq:attenuatedPostSelection}
    \vec{f_I^\gamma}\left(t+\tau\right)=
    \begin{pmatrix}
        f_{Ix} e^{-\frac{1}{2}\gamma \tau}\\
        f_{Iy} e^{-\frac{1}{2}\gamma \tau}\\
        f_{Iz}  e^{-\gamma \tau}
    \end{pmatrix}=e^{-\frac{1}{2}\gamma \tau}
    \begin{pmatrix}
        f_x \cos \left[\omega_a \left(t+\tau \right)\right]+f_y \sin \left[\omega_a \left(t+\tau \right)\right]\\
        f_y \cos \left[\omega_a \left(t+\tau \right)\right]-f_x \sin \left[\omega_a \left(t+\tau \right)\right]\\
        f_z \ e^{-\frac{1}{2}\gamma \tau}
    \end{pmatrix}.
\end{equation}
It is quite interesting to note that the consequences of the quantum system evolution under the full Lindbladian (\ref{eq:Lindblad_master_equation}) can be taken into account by specifying an attenuated post-selected state $\vec{f}_I^\gamma\left(t+\tau\right)$ that picks up the complete evolution after the weak interaction, including the free evolution and the dissipation. We see that the effect of dissipation for a long time $\tau$ is to drive the attenuated post-selected state to $\vec{f_I^\gamma}=0$. In that case, the density operator associated with the attenuated post-selected state becomes the maximally mixed state $\hat{\mathbb{1}}/2$. This explains why anomalousness disappears at long dissipation times. In that case, we also see that the weak value becomes real and converges towards the simple scalar product $\vec{i}\cdot\vec{n}_I$, which corresponds to the expectation value of the operator $\vec{n}_I\cdot\hat{\vec{\sigma}}$ in the initial state $\vec{i}$ expressed in terms of Bloch vectors (as shown in section \ref{sec:WVevol}). We should be careful, though, and note that the effect of dissipation is not strictly to post-select in $\vec{f_I^\gamma}$: indeed,  the formula giving the weak value with dissipation is not exactly the same as the weak value without dissipation with $\vec{f}_I$ replaced by $\Vec{f_I^\gamma}$. However, we see that the consequences of dissipation can always be anchored to the final state. In particular, if we set $\gamma=0$ to cancel the effects of dissipation, we recover the expression of a standard weak value of a Pauli operator \cite{mc2017,lbf2022}:
\begin{equation}\label{eq:WVNodissipation2lvlAnalytical}
   \lim_{\gamma \to 0} \sigma_{S,w}\left(\tau\right)=\frac{
    \vec{f}_I\cdot\vec{n}_I+\vec{i}\cdot\vec{n}_I + i \vec{f}_I\cdot \left(\vec{n}_I\cross\vec{i}\right)
    }{1+\vec{f}_I\cdot\vec{i}}.
\end{equation}
The structure of the expressions (\ref{eq:WVdissipation2lvlAnalytical}) and (\ref{eq:WVNodissipation2lvlAnalytical}) differ by the presence of two additional terms, proportional to the difference of the $z$ components of $\vec{f_I^\gamma}$ and $\vec{f}_I$, in the case of dissipation.

In practice, the effect of the system free evolution during the time $t+\tau$ is to rotate the post-selected vector (\ref{eq:RabiPostSelectedRotation}). In order to focus for the weak measurement specifically on the consequences of dissipation, we can choose to set, as done in section \ref{sec:WVevol} and \ref{sec:WVdissipation}, a constant post-selected vector in the interaction representation: $\vec{f}_I\left(t+\tau\right)=\left(f_{Ix0},f_{Iy0},f_{Iz0}\right)$ and, thus, $\vec{f_I^\gamma}\left(t+\tau\right)=\left(f_{Ix0}\, e^{-\frac{\gamma}{2}\tau},f_{Iy0}\, e^{-\frac{\gamma}{2}\tau},f_{Iz0}\, e^{-\gamma\tau}\right)$. In that case, the post-selection must depend on time $\tau$ in the Schrödinger representation, meaning that the experimentalist must rotate its post-selection choice as a function of the dissipation duration $\tau$:
\begin{equation}
    \vec{f}=
    \begin{pmatrix}
        f_{x} \\
        f_{y} \\
        f_{z}
    \end{pmatrix}=
    \begin{pmatrix}
        f_{Ix0} \cos \left[\omega_a \left(t+\tau \right)\right]-f_{Iy0} \sin \left[\omega_a \left(t+\tau \right)\right]\\
        f_{Iy0} \cos \left[\omega_a \left(t+\tau \right)\right]+f_{Ix0} \sin \left[\omega_a \left(t+\tau \right)\right]\\
        f_{Iz0}
    \end{pmatrix}.
\end{equation}
This time dependence in the Schrödinger representation requires the experimenter to choose a different post-selection depending on the time of post-selection.

Although we derived the expression of the weak value (\ref{eq:WVdissipation2lvlAnalytical}) in the context of the Rabi model, the expression is quite generally valid in the context of two-level systems. For an arbitrary observable of a two-level system undergoing dissipation as in (\ref{eq:dissipationIn2lvlMatrix}), we can write $\hat{A}_S=a\ \hat{\mathbb{1}} + b\ \vec{m}\cdot\hat{\vec{\sigma}}$. Then, for arbitrary initial and final states, the weak value expression is
\begin{equation}\label{eq:WVdissipation2lvlAnalytical-GeneralCase}
    A_{S,w}\left(\tau\right)=a + b \frac{
    \vec{f_I^\gamma}\cdot\vec{m}_I+\vec{i}\cdot\vec{m}_I \left(1+f_{Iz}^\gamma-f_{Iz}\right)+ i \vec{f_I^\gamma}\cdot \left(\vec{m}_I\cross\vec{i}\right)
    }{1+\vec{f_I^\gamma}\cdot\vec{i}+\left(f_{Iz}^\gamma-f_{Iz}\right)}.
\end{equation}
with $\vec{f}_I$ the Bloch vector representing the post-selected state at time $t+\tau$ in the interaction picture and $\vec{f_I^\gamma}$, the attenuated post-selected vector, constructed from $\vec{f}_I$ as in the left-hand side of (\ref{eq:attenuatedPostSelection}), while $\vec{m}_I$ is the Bloch vector associated to $\hat{A}_{SI}\!\left(t/2\right)$ in the interaction picture. Within the weak value, the attenuated post-selection vector $\vec{f_I^\gamma}$ contains the information on the dissipation of the system after the weak interaction, which can be analyzed through the related weak measurement shifts.
\subsection{Rotating-wave approximation}
 The standard weak measurement approximation enables expressing the final meter state as a function of the weak value of the weakly measured Hermitian operator in the pre- and post-selected quantum system. In the general Rabi model, this requires $\omega_a t \ll 1$ and $\omega_f t \ll 1$, meaning that the free evolution can be evaluated to first order during the short interaction time $t$. However, this assumption may not always be warranted. For example, the transit time of an atom passing through a cavity may be several orders of magnitude longer than the oscillation period of the cavity field. In that case, considering a weakly coupled atom--cavity system near resonance, the atomic Hamiltonian will also generate many oscillations during the interaction as $\left\vert\Delta\right\vert \ll \omega_a + \omega_f = 2 \omega$ with $\Delta= \omega_a - \omega_f$. In order to make progress, we thus have to come back to the atom-field joint state after the weak interaction given by (\ref{eq:evolution_rho_interaction_picture_after_short_time}), knowing that the approximation (\ref{eq:interactionHamiltonianApprox}) is now invalid. Then, we have to evaluate the integral of the interaction Hamiltonian in the interaction picture:
\begin{eqnarray}
     \int_0^t \hat{V}\left(t'\right) dt' &=& \hbar g_{\mathrm{int}} \int_0^t \hat{\sigma}_{xI}\left(t'\right)\otimes \left[\hat{a}_I^\dagger\left(t'\right)+\hat{a}_I\left(t'\right)\right]dt' \label{eq:integralJC1}\\
     &=&\hbar g_{\mathrm{int}} \int_0^t  \left[\hat{\sigma}_{+} e^{i \omega_a t^\prime}+\hat{\sigma}_{-} e^{-i \omega_a t^\prime}\right]\otimes \left[\hat{a}^\dagger e^{i \omega_f t^\prime}+\hat{a}e^{-i \omega_f t^\prime}\right]dt'\label{eq:integralJC2}\\
     &\approx&\hbar g_{\mathrm{int}} \int_0^t \left[e^{i \left(\omega_a - \omega_f\right) t^\prime} \hat{\sigma}_{+} \otimes \hat{a}+e^{-i \left(\omega_a - \omega_f\right) t^\prime} \hat{\sigma}_{-} \otimes \hat{a}^\dagger\right]dt' \label{eq:integralJC3}\\
     &=&\hbar g_{\mathrm{int}} \frac{\sin{\Delta t}}{\Delta}\left[e^{i \Delta t /2} \hat{\sigma}_{+} \otimes \hat{a}+e^{- i \Delta t /2} \hat{\sigma}_{-} \otimes \hat{a}^\dagger\right] \label{eq:integralJC4}\\
     &\approx&\hbar g_{\mathrm{int}} t\left[e^{i \Delta t /2} \hat{\sigma}_{+} \otimes \hat{a}+e^{- i \Delta t /2} \hat{\sigma}_{-} \otimes \hat{a}^\dagger\right], \label{eq:integralJC5}
 \end{eqnarray}
where we used the fact that the atom and field Hamiltonians commute (\ref{eq:integralJC1}); we developed $\hat{\sigma}_x=\hat{\sigma}_{+}+\hat{\sigma}_{-}$ (\ref{eq:integralJC2}); we neglected the fast oscillating terms $\hat{\sigma}_{-}\hat{a}^\dagger e^{2 i \omega}$ and $\hat{\sigma}_{+}\hat{a} e^{-2 i \omega}$ with respect to the slow ones (\ref{eq:integralJC3}), as they will contribute a term in $1/(2\omega)$ in (\ref{eq:integralJC4}); and where we expanded $\sin({\Delta t})$ to first order in (\ref{eq:integralJC5}), assuming that the interaction time is short with respect to the detuning $\Delta t \ll 1$.
In practice, the rotating-wave approximation that we carried out leads to the conservation of the excitation number, by neglecting the non-resonant terms originating from $\hat{H}_{\mathrm{int}}$ in Eq.~(\ref{eq:hamiltonians_rabi_model}). Note that the rotating-wave approximation also requires the condition $\omega_f\gg g_{\mathrm{int}}$, which is met in most, if not all, atomic cavity QED experiments. 

We observe that the effective interaction Hamiltonian in the weak measurement and rotating-wave approximations comprises two terms built from non-Hermitian operators. In the first one, the raising operator $\hat{\sigma}_{+}$ of the system is coupled to the annihilation operator $\hat{a}$ of the meter, while, in the second term, the lowering operator $\hat{\sigma}_{-}$ of the system is coupled to the creation operator $\hat{a}^\dagger$ of the meter. The sum of the two coupling terms is nevertheless a Hermitian operator. Starting from the effective interaction Hamiltonian $V_I\left(t/2\right)$ given by (\ref{eq:integralJC5}), by following the exact same theoretical developments performed in section \ref{sec:generalTheory}, beginning from equation (\ref{eq:evolution_rho_interaction_picture_after_dissipation}), we deduce immediately the equivalent of equation (\ref{eq:expectation_value_L_dissipation_WithDenom})
\begin{equation}
\label{eq:JaynesCummingWM-general}
\langle\hat{L}\rangle_f=\frac{\langle\hat{L}_I\rangle_0+2 g_{\mathrm{int}} t\,\text{Im}\left[e^{i \Delta t /2} \sigma_{+,w}\left(\tau\right)\langle\hat{L}_I\hat{a}\rangle_0\right]+2 g_{\mathrm{int}} t\,\text{Im}\left[e^{-i \Delta t /2} \sigma_{-,w}\left(\tau\right)\langle\hat{L}_I\hat{a}^\dagger\rangle_0\right]}{1+2 g_{\mathrm{int}} t \Im \left[e^{i \Delta t /2}\sigma_{+,w}\left(\tau\right) \langle\hat{a}\rangle_0\right]+2 g_{\mathrm{int}} t \Im \left[e^{-i \Delta t /2}\sigma_{-,w}\left(\tau\right) \langle\hat{a}^\dagger\rangle_0\right]},
\end{equation}
with $\hat{L}_I\!\left(t+\tau\right)$ and the weak values 
\begin{equation}\label{eq:WVraising-lowering}
 \sigma_{\pm,w}\left(\tau\right)=\frac{\text{Tr}\left[\hat{\sigma}_{fI}\!\left(t+\tau\right)\ e^{\mathcal{D}\tau}\left(\hat{\sigma}_{\pm}\hat{\sigma}_i\right)\right]}{\text{Tr}\left[\hat{\sigma}_{fI}\!\left(t+\tau\right)\ e^{\mathcal{D}\tau}\left(\hat{\sigma}_i\right)\right]}.
\end{equation}
We would like to point out that the operators $\hat{\sigma}_\pm$ appear in the Schrödinger representation, as their time dependence has been explicitly factored out in (\ref{eq:integralJC5}) and (\ref{eq:JaynesCummingWM-general}). The meter shift exhibits now two contributions, one from each coupling term. Very interestingly, the weak measurement result depends on the weak values of the non-Hermitian raising and lowering operators $\hat{\sigma}_{\pm}$ of the system.

To determine the weak values, we consider measuring the two meter quadratures  $\hat{Q}_I\!\left(t+\tau\right)$ and $\hat{P}_I\!\left(t+\tau\right)$. Assuming that the initial meter state is an energy eigenstate $\ketbra{n}{n}$ or a thermal state, we have $\langle\hat{a}\rangle_0=\langle\hat{a}^\dagger\rangle_0=0$ in the denominator, as well as  $\langle\hat{L}_I\rangle_0=0$ in the numerator. We get thus the simple expressions
\begin{eqnarray}
\label{eq:JaynesCummingWM-numberthermalstate}
\langle\hat{Q}\rangle_f=2 g_{\mathrm{int}} t\sqrt{\frac{2}{\hbar\omega_f}}\text{Im}\left[e^{i \Delta t /2+\omega_f\left(t+\tau\right)} \sigma_{+,w}\left(\tau\right)\langle\hat{a}^\dagger\hat{a}\rangle_0+e^{-i \Delta t /2-i \omega_f\left(t+\tau\right)} \sigma_{-,w}\left(\tau\right)\langle\hat{a}\hat{a}^\dagger\rangle_0\right],\\
\langle\hat{P}\rangle_f=2 g_{\mathrm{int}} t\sqrt{\frac{2 \omega_f}{\hbar}}\text{Re}\left[e^{i \Delta t /2+\omega_f\left(t+\tau\right)} \sigma_{+,w}\left(\tau\right)\langle\hat{a}^\dagger\hat{a}\rangle_0-e^{-i \Delta t /2-i \omega_f\left(t+\tau\right)} \sigma_{-,w}\left(\tau\right)\langle\hat{a}\hat{a}^\dagger\rangle_0\right],
\end{eqnarray}
with $\langle\hat{a}^\dagger\hat{a}\rangle_0=n$ and $\langle\hat{a}\hat{a}^\dagger\rangle_0=n+1$, where $n$ labels the energy level in an energy eigenstate, or corresponds to the average energy level $n_{\mathrm{eq}}=\left[e^{\hbar \omega_f / \left(k_B T\right)}-1\right]^{-1}$ in a thermal-equilibrium state of the meter. We thus see that there are generally two contributions to the observed shifts, arising from two weak values.

The simplest case, yet fascinating, corresponds to a meter initially in the vacuum state $\ketbra{0}{0}$, so that only the weak value of $\hat{\sigma}_{-}$ contributes to the shift since $\langle\hat{a}^\dagger\hat{a}\rangle_0=0$. In that case, the weak measurement expectation values are given by
\begin{eqnarray}
\langle\hat{Q}\rangle_{f}&=&2 g_{\mathrm{int}} t \sqrt{\frac{\hbar}{2\omega_f}} \left\vert \sigma_{-,w}\left(\tau\right) \right\vert \sin{\left[\varphi_w(\tau)-\Delta t/2-\omega_f (t+\tau)\right]}, \\
\langle\hat{P}\rangle_{f}&=&-2 g_{\mathrm{int}} t \sqrt{\frac{\hbar \omega_f}{2}} \left\vert \sigma_{-,w}\left(\tau\right) \right\vert \cos{\left[\varphi_w(\tau)-\Delta t/2-\omega_f (t+\tau)\right]},
\end{eqnarray}
where $\varphi_w(\tau)$ is the argument of the weak value of $\hat{\sigma}_{-}$ (\ref{eq:WVraising-lowering}). These shifts should be compared to the results obtained before in the Rabi model (\ref{eq:QRabiQuadrature}) and (\ref{eq:PRabiQuadrature}), which are functions of the weak value of the operator $\hat\sigma_{xI}(t/2)$. In a weak measurement performed within the validity of the Jaynes-Cumming model with a meter initially in the vacuum state, the meter shifts depend on the single weak value of the lowering operator of the system, a non-Hermitian operator. This approach enables using the weak value approximation for interaction times $t$ such that $\omega_f t \gg 1$ and $\omega_a t \gg 1$, on the condition that the interaction time remains short with respect to the frequency detuning $\Delta t \ll 1$. In these circumstances, it is possible to investigate the dissipation dynamics of the system through the weak value of the non-Hermitian operator $\hat{\sigma}_{-}$.

The general formula (\ref{eq:WVdissipation2lvlAnalytical-GeneralCase}) for a two-level weak value affected by dissipation enables expressing the weak value of the non-Hermitian operators $\hat{\sigma}_\pm$ straightforwardly. By setting $a=b=1$ and using a complex three-dimensional vector $\vec{m}_\pm=(1, \pm i,0)/2$ because $2\hat{\sigma}_{\pm}=\sigma_x\pm i \sigma_y$, we obtain
\begin{equation}\label{eq:WVdissipation2lvlAnalytical-NonHermitian}
    \sigma_{\pm,w}\left(\tau\right)=\frac{
    \vec{f_I^\gamma}\cdot\vec{m}_{\pm}+\vec{i}\cdot\vec{m}_\pm \left(1+f_{Iz}^\gamma-f_{Iz}\right)+ i \vec{f_I^\gamma}\cdot \left(\vec{m}_\pm\cross\vec{i}\right)
    }{1+\vec{f_I^\gamma}\cdot\vec{i}+\left(f_{Iz}^\gamma-f_{Iz}\right)},
\end{equation}
which becomes
\begin{eqnarray}
     \sigma_{-,w}\left(\tau\right)&\mkern-9mu=&\mkern-9mu\frac{i_x \left(1-f_{Iz}\right)+f_{Ix}^\gamma \left(1+i_z\right)-i \left[i_y\left(1-f_{Iz}\right)+f_{Iy}^\gamma \left(1+i_z\right)\right]
    }{2\left[1+\vec{f_I^\gamma}\cdot\vec{i}+\left(f_{Iz}^\gamma-f_{Iz}\right)\right]},\\
    \sigma_{+,w}\left(\tau\right)&\mkern-9mu=&\mkern-9mu\frac{i_x \left(1-f_{Iz}+2 f_{Iz}^\gamma\right)+f_{Ix}^\gamma \left(1-i_z\right)+i \left[i_y\left(1-f_{Iz}+2 f_{Iz}^\gamma\right)+f_{Iy}^\gamma \left(1-i_z\right)\right]
    }{2\left[1+\vec{f_I^\gamma}\cdot\vec{i}+\left(f_{Iz}^\gamma-f_{Iz}\right)\right]}.
\end{eqnarray}
These weak values appear relatively simple and symmetric,  with only a few Bloch vector components involved in their numerators. These expressions are helpful to choose appropriate pre- and post-selected states $\Vec{i}$ and $\Vec{f}$ to probe the dissipation dynamics that are revealed through the contributions of the $\vec{f_I^\gamma}$ vectors to the meter shifts.

\section{Exploiting weak values in dissipative systems}\label{sec:WVdissipation}

Dissipation has a negative impact on the amplification properties of the weak value. In systems with non-degenerate energy levels, weak values cannot provide amplification when the dissipation time is long. In general, the longer the dissipation time, the weaker the amplification of the weak value. Nevertheless, measuring the weak value at short dissipation times can still provide valuable information about the evolution of the system. For example, the rate of change of the weak value could be used to determine the dissipation rate, allowing shorter measurement times than those required by traditional methods.

It is often interesting to detect non-markovianity in the evolution of a quantum system, which may not always be straightforward. For certain non-Markovian evolutions, we show it is possible to differentiate Markovian dissipation from non-Markovian dissipation by analysing the evolution of the weak value. This result demonstrates the potential of weak measurement theory in identifying non-Markovian dynamics.

\subsection{Effective amplification of the dissipation rate}\label{subsec:WVdissipation-Markovian}

In this section, we show that the dissipation rate can, in principle, be accurately determined from the evolution of weak values in a short interval of time. To begin with, let us examine the atom evolution in the dissipative regime in the interaction picture,
\begin{equation}\label{eq:Lindblad_evolution_section5}
\dot{\hat{\sigma}}_{SI}=\mathcal{D}\left(\hat{\sigma}_{SI}\right)=\gamma\left[\hat{\sigma}_{-}\hat{\sigma}_{SI}\hat{\sigma}_{+}-\frac{1}{2}\left(\hat{\sigma}_{+}\hat{\sigma}_{-}\hat{\sigma}_{SI}+\hat{\sigma}_{SI}\hat{\sigma}_{+}\hat{\sigma}_{-}\right)\right],
\end{equation}
where dissipation is assumed to be Markovian. 

For this illustration, we will consider a weak measurement of the $\hat{\sigma}_x$ operator, i.e. $\hat{A}_{SI}\left(t/2\right)=\hat{\sigma}_x$ in the weak value expression (\ref{eq:weak_value_with_dissipation}) and, equivalently, $\vec{n}_I=\left(1,0,0\right)$ in (\ref{eq:WVdissipation2lvlAnalytical}). We also take a post-selected state that is constant in the interaction representation, namely $\hat{\sigma}_{fI}\left(t+\tau\right)=\hat\sigma_{fI0}$. As a result, we focus our attention exclusively on measuring the dissipation, without probing the free Hamiltonian evolution. Since the dissipation time is short, we can expand the weak value of the spin operator $\hat{\sigma}_x$ in a Taylor series up to the first order in $\tau$,
\begin{eqnarray}
\label{eq:weak_value_Taylor}
\sigma_{x,w}\left(\tau\right)&=&\frac{\text{Tr}\left[\hat{\sigma}_{fI0}\,e^{\mathcal{D}\tau}\left(\hat{\sigma}_x\hat{\sigma}_i\right)\right]}{\text{Tr}\left[\hat{\sigma}_{fI0}\,e^{\mathcal{D}\tau}\left(\hat{\sigma}_i\right)\right]}\\ \nonumber
&\approx&\frac{\text{Tr}\left[\hat{\sigma}_{fI0}\,\hat{\sigma}_x\hat{\sigma}_i\right]}{\text{Tr}\left[\hat{\sigma}_{fI0}\,\hat{\sigma}_i\right]}+\tau \left.\frac{d}{d \tau}\left(\frac{\text{Tr}\left[\hat{\sigma}_{fI0}\,e^{\mathcal{D}\tau}\left(\hat{\sigma}_x\hat{\sigma}_i\right)\right]}{\text{Tr}\left[\hat{\sigma}_{fI0}\,e^{\mathcal{D}\tau}\left(\hat{\sigma}_i\right)\right]}\right) \right\vert_{\tau=0},
\end{eqnarray}
where the derivative can be computed using Eq.~(\ref{eq:Lindblad_evolution_section5}). Let $\epsilon \ll 1$ be a small number, which we will show is inversely proportional to the amplification of the weak value. When the initial and the post-selected atomic states are chosen, to first order in $\epsilon$, as
\begin{eqnarray}
\label{equation:pre_and_post_selected_states_development_non_markovian_and_dissipation}
\ket{\psi_i}&=&-\text{sign}\left(\epsilon\right)\ket{g}+\frac{|\epsilon|}{2}\ket{e},\\ \nonumber
\ket{\psi_{fI0}}&=&\frac{1}{\sqrt{2}}\left(\epsilon\ket{g}+\left(1-i\right)\ket{e}\right),
\end{eqnarray}
with $\hat{\sigma}_{fI0}=\ketbra{\psi_{fI0}}{\psi_{fI0}}$, then the probability of post-selection (i.e.\ the denominator of Eq.~(\ref{eq:weak_value_Taylor})) is $p=\epsilon^2/4$ for $\tau=0$
and the weak value at first order in $\gamma\tau$ and $\epsilon$ is given by\footnote{A few technical considerations on the joint series developments are provided in Appendix \ref{appendix:Markovian}}
\begin{equation}
\label{eq:weakvalueapproxMarkov}
\sigma_{x,w}\approx\frac{\tau \gamma}{\epsilon}+i\frac{2-\tau\gamma}{\epsilon}.
\end{equation}
We see that for a given time $\tau$, the dissipation rate $\gamma$ is effectively amplified by the small parameter $\epsilon$ in the denominator of the real part of the weak value. At $\tau=0$, the weak value is purely imaginary. However, when dissipation is introduced ($\tau\neq 0$), the weak value becomes a complex number that varies linearly with time, $\tau\gamma$. In principle, the dissipation rate can be extracted by measuring the real part of the weak value at different short times. This approach takes advantage of the weak value amplification, proportional to $\frac{1}{\epsilon}$, which opens up the possibility of using much shorter dissipation times. This is especially relevant for experiments where it may be difficult to obtain long dissipation times. In particular, by resorting to amplification, we reduce the measurement duration, which may be helpful if the meter undergoes some dissipation in practice (contrary to our model assumptions). We note that it is also possible to extract the dissipation rate by measuring the weak value for several values of the small parameter $\epsilon$ for a fixed duration $\tau$ of the dissipation, if this proves more convenient.

\subsection{Revealing non-markovianity}\label{subsec:WVdissipation-nonMarkovian}
In this section, we show that weak values can provide a valuable tool to distinguish between certain Markovian and non-Markovian dynamics. More specifically, we consider here a modified experimental setting in which a two-level atom undergoes a weak interaction in a cavity and then enters a second leaky cavity. Inside the second cavity, the atom is coupled to a single cavity mode which is itself coupled to the bosonic bath in vacuum associated with the field outside the cavity. The dynamics of the atom in this cavity is described by the Jaynes-Cummings model on resonance, where the Hamiltonian of the atom is proportional to $\hat{\sigma}_z$, as explained in section \ref{sec:RabiModel}. In this model, the non-Markovian dissipator is given by \cite{breuer2002theory}
\begin{equation}\label{eq:dissipatorNonMarkovian}
\mathcal{D}\left(\hat\sigma_{SI}\right)=\gamma\left(\tau\right)\left[\hat{\sigma}_{-}\hat\sigma_{SI}\hat{\sigma}_{+}-\frac{1}{2}\left(\hat{\sigma}_{+}\hat{\sigma}_{-}\hat\sigma_{SI}+\hat\sigma_{SI}\hat{\sigma}_{+}\hat{\sigma}_{-}\right)\right],
\end{equation} 
with the time-dependent dissipation rate
\begin{equation}\label{eq:defgammaoftau}
\gamma\left(\tau\right)=\frac{2\gamma_0\lambda\sinh{(d\frac{\tau}{2})}}{d\cosh{(d\frac{\tau}{2})}+\lambda\sinh{(d\frac{\tau}{2})}},
\end{equation}
where $d=\sqrt{\lambda^2-2\gamma_0\lambda}$, $\lambda$ defines the spectral width of the coupling, in other words, the inverse of the bath correlation time, and $\gamma_0$ is
the typical atomic decay rate in the Markovian limit. 
The nature of the parameter $d$ in the Jaynes-Cummings model on resonance varies based on specific conditions \cite{breuer2002theory}. In cases of moderate or weak coupling ($\lambda^2 > 2\gamma_0\lambda$), $d$ takes on a real value, resulting in the absence of oscillations in the system dynamics. On the other hand, under strong coupling conditions ($\lambda^2 < 2\gamma_0\lambda$), $d$ is imaginary and the characteristic oscillatory behavior emerges. The dynamics generated by this model is non-Markovian \cite{breuer2002theory}, which means that the evolution of the system at each time step depends on its past evolution, not just its present state. If we select the pre- and post-selected states as in the previous Markovian case (Eq.~(\ref{equation:pre_and_post_selected_states_development_non_markovian_and_dissipation})), the weak value of $\hat\sigma_x$ at second order in $\tau$ and first order in $\epsilon$ is
\begin{equation}
\label{eq:non_markovian}
\sigma_{x,w}\left(\tau\right)\approx \frac{\lambda\tau^2 \gamma_0}{2\epsilon}+i\frac{4-\lambda\tau^2\gamma_0}{2\epsilon},
\end{equation}
where the exact analytical solution of the non-Markovian evolution of the weak value as a function of $\tau$ was used and expanded in Taylor series for small values of $\tau$ (see Appendix \ref{appendix:NonMarkovian} for details). The above expression is valid if
\begin{equation}
\lambda\tau \ll 1 \quad\mathrm{and}\quad \gamma_0\tau\ll 1.
\end{equation}
The non-Markovian expression (\ref{eq:non_markovian}) can be recovered from the Markovian case if we replace $\gamma$ in Eq.~(\ref{eq:weakvalueapproxMarkov}) by $\frac{1}{2}\gamma_0 \lambda \tau$, which corresponds to the series expansion of $\gamma(\tau)$ to first order in $\lambda \tau$ in Eq.~(\ref{eq:defgammaoftau}).
We can see from Eq.~(\ref{eq:non_markovian}) that the weak value exhibits a quadratic evolution with respect to $\tau$, with no linear term in $\tau$. This behavior is characteristic of non-Markovian dynamics described by the above dissipator, provided that $\gamma\left(\tau=0\right)=0$. To consider the transition from non-Markovian to Markovian dynamics in this system, we highlight that the Jaynes-Cummings model becomes Markovian as $\lambda$ approaches infinity, with $\lim_{\lambda \to \infty} \gamma\left(\tau\right)=\gamma_0$. Consequently, the series expansion leading to (\ref{eq:non_markovian}) becomes invalid.
Hence, the weak value (\ref{eq:non_markovian}) corresponds to the strongly non-Markovian case.

By comparing the evolution of the weak value as a function of $\tau$ at short dissipation times, it is possible to distinguish between Markovian and non-Markovian dynamics. Identically to the Markovian case,  amplification occurs when $\epsilon\ll 1$. However, in strongly non-Markovian systems, the weak value is proportional to $\tau^2$ instead of $\tau$. The real part of the weak value (\ref{eq:non_markovian}) is null at $\tau=0$, making detection of the quadratic dependence easier.
By examining the relationship between Eq.~(\ref{eq:weakvalueapproxMarkov}) and Eq.~(\ref{eq:non_markovian}), we can obtain valuable information about $\lambda$, especially when $\gamma_0=\gamma$. By incorporating a leaky cavity and comparing the weak value evolution to its free space counterpart (without the cavity), we can extract information about the leaky cavity parameters.

In summary, we have shown that the weak value at short times provides valuable information about dissipation in a system through an amplified decay rate, the effects of which are felt over a shorter evolution time in practice.

\section{Conclusions}
In this paper, we outlined a few benefits that weak measurements and, in particular, weak values bring to the study of open quantum systems. We have shown that dissipation impedes the amplification produced by weak values at long times, unless the system's ground state is degenerate. More specifically, we have considered a pre-selection, weak interaction, dissipation, and post-selection scheme. In non-degenerate cases, we have shown that the limit of the weak value at infinite dissipation time is the expectation value of the operator in the initial state, so that dissipation suppresses the anomalous character associated with weak values and prevents amplification. However, in systems with a degenerate ground state, amplification can still occur at very long timescales, as we have illustrated with a specific system. 

Dissipation-induced decay to a degenerate ground state can even increase the anomalous properties of the weak value until it stabilizes. This is possible because not all quantum properties are lost through dissipation and the final state reached after long dissipation times can still be a combination of different ground states that retain some coherence. In particular, observing anomalous weak values at long dissipation time requires that the unperturbed initial state and its perturbation provoked by the weak interaction evolve differently under dissipation.

In addition, weak values can be used to measure various properties of the evolution of open quantum systems. For example, by choosing appropriate pre- and post-selected states, we can extract information about the dissipation rate through the weak value at short dissipation times, in the weak value amplification regime. This is particularly helpful if experimental constraints require a short measurement duration of each quantum system. 
We also explained how it is possible to distinguish between Markovian and non-Markovian evolutions by measuring with our scheme the growth rate of the weak value. For Markovian evolutions, the weak value always increases linearly with time. On the other hand, in the strongly non-Markovian regime, the weak value at small times increases quadratically with time. Consequently, a measure of the growth rate of the weak value is sufficient to distinguish the two contrasting cases. 
Besides, we observed that using a cavity mode as the meter of a weak measurement of an atom's internal degrees of freedom may yield meter measurement results that depend on weak values of the non-Hermitian raising and lowering operators of the atom.
These applications demonstrate the usefulness of weak values in open quantum systems. We hope that this work will set the stage for other applications connecting weak values and open quantum systems.

\begin{backmatter}
\bmsection{Acknowledgments}
Y.C. is a Research Associate of the Fund for Scientific Research F.R.S.-FNRS. This research was supported by the Action de Recherche Concertée WeaM at the University of Namur (19/23-001). L.B.F. acknowledges WeaM for funding and the University of Liège for hosting her regularly during the course of this work.
\end{backmatter}

\appendix
\section{Appendix}
\subsection{Expression of the jump operators for the degenerate ground state study}\label{appendix:expression_l_q}
The expression of the jump operators $\hat{L}_q$ used in Eq.~(\ref{eq:L_I}) are
\begin{eqnarray}
    \hat{L}_0&=&\sqrt{\tfrac{2}{3}}\,\ket{J_g,-\tfrac{1}{2}}\bra{J_e,-\tfrac{1}{2}}+\sqrt{\tfrac{2}{3}}\,\ket{J_g,\tfrac{1}{2}}\bra{J_e,\tfrac{1}{2}}, \\ \nonumber
    \hat{L}_{-}&=&\ket{J_g,-\tfrac{1}{2}}\bra{J_e,-\tfrac{3}{2}}+\tfrac{1}{\sqrt{3}}\,\ket{J_g,\tfrac{1}{2}}\bra{J_e,-\tfrac{1}{2}}, \\ \nonumber
    \hat{L}_{+}&=&\tfrac{1}{\sqrt{3}}\,\ket{J_g,-\tfrac{1}{2}}\bra{J_e,\tfrac{1}{2}}+\ket{J_g,\tfrac{1}{2}}\bra{J_e,\tfrac{3}{2}}.
\end{eqnarray}

\subsection{Extracting the Weak value from the meter measurements in the Rabi model}\label{appendix:WVfromWM}
To shorten the formulas, we denote the anti-commutator averages by $ACP=\langle\left\{\hat{P}_I\left(t+\tau)\right),\hat{N}_I\left(t/2\right)\right\}\rangle_0 $ and  $ACQ=\langle\left\{\hat{Q}_I\left(t+\tau)\right),\hat{N}_I\left(t/2\right)\right\}\rangle_0 $; the commutator averages by $CP=\langle\left[\hat{P}_I\left(t+\tau)\right),\hat{N}_I\left(t/2\right)\right]\rangle_0 $ and  $CQ=\langle\left[\hat{Q}_I\left(t+\tau)\right),\hat{N}_I\left(t/2\right)\right\rangle_0 $; the meter operators averages in the initial meter state by $N0=\langle\hat{N}_I\left(t/2\right)\rangle_0$, $Q0=\langle\hat{Q}_I\left(t+\tau\right)\rangle_0$, and $P0=\langle\hat{P}_I\left(t+\tau\right)\rangle_0$. Then, the expression of the weak value in terms of the measurement results, namely the averages $\langle\hat{Q}\rangle_f$ and $\langle\hat{P}\rangle_f$ of the two meter quadratures, is
\begin{eqnarray}
\label{eq:WVinversion}
\text{Re}A_{S,w}&\mkern-9mu=&\mkern-9mu\frac{i}{gt} \frac{ACQ \left(\langle\hat{P}\rangle_f-P0\right)-ACP \left(\langle\hat{Q}\rangle_f-Q0\right)+2 N0 \left(P0\ \langle\hat{Q}\rangle_f-\langle\hat{P}\rangle_f\ Q0\right)}{ACQ\ CP-ACP\ CQ+2 N0 \left(CQ\ \langle\hat{P}\rangle_f - CP\ \langle\hat{Q}\rangle_f\right)},\\
\text{Im}A_{S,w}&\mkern-9mu=&\mkern-9mu-\frac{1}{gt} \frac{CQ \left(\langle\hat{P}\rangle_f-P0\right)-CP \left(\langle\hat{Q}\rangle_f-Q0\right)}{ACQ\ CP-ACP\ CQ+2 N0 \left(CQ\ \langle\hat{P}\rangle_f - CP\ \langle\hat{Q}\rangle_f\right)},
\end{eqnarray}
which is valid for an arbitrary initial meter state. These expressions are obtained by inverting (\ref{eq:expectation_value_L_RealCommutator-ImaginaryAntiCommutator_WithDenominatorAdded}) using both quadratures. In order to retrieve the weak value from the quadrature measurements, we see that it is preferable to choose an initial meter state verifying $N0=0$, as well as $P0=Q0=0$.

\subsection{Series expansion of the weak value (Markovian case)}\label{appendix:Markovian}
We discuss here a few technical issues related to the series expansion of the weak value (\ref{eq:weakvalueapproxMarkov}) in section \ref{subsec:WVdissipation-Markovian}. Let us consider the denominator of the two-level system weak value (\ref{eq:WVdissipation2lvlAnalytical}):
\begin{equation}\label{eq:WVdissipation2lvlAnalytical-denominator}
    1+\left(f_{Ix} i_x + f_{Iy} i_y \right) e^{-\frac{1}{2}\gamma \tau} + f_{Iz} \left(1+ i_z\right) e^{-\gamma\tau}-f_{Iz}.
\end{equation}
Since we would like to exploit amplification, we require that the pre- and post-selected states are nearly orthogonal $\left\vert\braket{\psi_{fI}}{\psi_i}\right\vert^2=\epsilon^{\prime 2}$ for a small parameter $\epsilon^\prime$. In terms of Bloch vectors, we have thus $\vec{f}_{I}\cdot\vec{i}=-1+2\epsilon^{\prime 2}$. We can then eliminate the $x$ and $y$ components of the Bloch vectors in (\ref{eq:WVdissipation2lvlAnalytical-denominator}): 
\begin{equation}\label{eq:WVdissipation2lvlAnalytical-denominator-onlyZ2}
    1-e^{-\frac{1}{2}\gamma \tau} + 2 \epsilon^{\prime 2} e^{-\frac{1}{2}\gamma \tau} + f_{Iz} \left(e^{-\gamma \tau} -1\right) + f_{Iz} i_z  \left(e^{-\gamma\tau}-e^{-\frac{1}{2}\gamma \tau}\right).
\end{equation}
Now, a first-order series expansion in $\gamma \tau$ yields
\begin{equation}\label{eq:WVdissipation2lvlAnalytical-denominator-onlyZ2-series}
     2 \epsilon^{\prime 2} + \frac{1}{2} \gamma \tau \left(1-2\epsilon^{\prime 2} - 2 f_{Iz} - f_{Iz} i_z \right) + \mathcal{O}\left(\gamma^2 \tau^2\right).
\end{equation}
This shows that we should be careful when making a series expansion of the full weak value with respect to $\gamma \tau$, as in (\ref{eq:weak_value_Taylor}), because we have to ensure that the first-order term in $\gamma \tau$ in the weak value denominator is smaller than the small term $\epsilon^{\prime 2}$ linked to the low post-selection probability. This would require $\gamma \tau \ll \epsilon^{\prime 2}$ in general (for example if choosing $f_{Iz}=0$), which could be an inconvenience as this imposes a bound on the amplification yield. However, if we ensure that the factor multiplying $\gamma \tau$ in (\ref{eq:WVdissipation2lvlAnalytical-denominator-onlyZ2-series}) is proportional to $\epsilon^{\prime 2}$, then the series expansion in $\gamma \tau$ is valid as long as $\gamma \tau \ll 1$, without constraining the amplification. Indeed, if we take the pre-and post-selected states with small components along $x$ and $y$ that are of the order of $\epsilon^\prime$, then, to second-order in $\epsilon^\prime$, we can write $f_{Iz}\approx 1- \alpha \epsilon^{\prime 2}$ and $i_{z}\approx -1+ \beta \epsilon^{\prime 2}$ (exchanging the roles of $f_{Iz}$ and $i_z$ works as well), where $\alpha$ and $\beta$ are unimportant proportionality constants related to the state normalization. This situation corresponds to pre- and post-selected states that nearly coincide with the basis states $\ket{e}$ and $\ket{g}$, respectively. Then, (\ref{eq:WVdissipation2lvlAnalytical-denominator-onlyZ2-series}) becomes
\begin{equation}\label{eq:WVdissipation2lvlAnalytical-denominator-onlyZ2-seriesfinal}
     2 \epsilon^{\prime 2} \left[1- \frac{1}{4} \gamma \tau \left(2-\alpha + \beta - \alpha \beta \epsilon^{\prime 2}\right)+ \mathcal{O}\left(\gamma^2 \tau^2\right)\right].
\end{equation}
The factorization of $\epsilon^{\prime 2}$ does not depend on the series expansion in $\gamma \tau$ and occurs also in (\ref{eq:WVdissipation2lvlAnalytical-denominator-onlyZ2}). With this choice of pre- and post-selected states, we obtain an amplification in $1/\epsilon^\prime$, while the series expansion to first order in $\gamma \tau$ is valid.

\subsection{Computation of the weak value (non-Markovian case)}\label{appendix:NonMarkovian}

With the non-Markovian dynamics specified in section \ref{subsec:WVdissipation-nonMarkovian} with the dissipator $\mathcal{D}$ defined in (\ref{eq:dissipatorNonMarkovian}) and the time-dependant dissipation rate given in (\ref{eq:defgammaoftau}), an arbitrary matrix $C=c_{gg}\ketbra{g}{g}+c_{ge}\ketbra{g}{e}+c_{eg}\ketbra{e}{g}+c_{ee}\ketbra{e}{e}$ becomes
\begin{equation}\label{eq:dissipationIn2lvlMatrix-NonMarkovian}
    e^{\mathcal{D}\tau}\left(C\right)=
    \begin{pmatrix}
          c_{ee}\ \Gamma^2 & c_{eg}\ \Gamma \\
           c_{ge}\ \Gamma & c_{gg}+c_{ee}\left(1-\Gamma^2\right)
    \end{pmatrix},
\end{equation}
where
\begin{equation}
    \Gamma=
    \frac{\sqrt{\lambda-\gamma_0-\gamma_0\cosh \left[\tau  \sqrt{\lambda  (\lambda -2 \gamma_0)}\right]}}{\sqrt{\lambda -2 \gamma_0}}
    \exp\left\{-\frac{\lambda  \tau }{2}+\tanh^{-1}\left[\frac{\sqrt{\lambda } \tanh \left(\frac{1}{2} \sqrt{\lambda } \tau  \sqrt{\lambda -2 \gamma_0}\right)}{\sqrt{\lambda -2 \gamma_0}}\right]\right\}.
\end{equation}
This evolution should be compared with the Markovian result (\ref{eq:dissipationIn2lvlMatrix}), which exhibits an exponential decay. Considering the similar structure of the solutions (\ref{eq:dissipationIn2lvlMatrix}) and (\ref{eq:dissipationIn2lvlMatrix-NonMarkovian}), the two-level weak value in the non-Markovian case is identical to (\ref{eq:WVdissipation2lvlAnalytical}) and (\ref{eq:WVdissipation2lvlAnalytical-GeneralCase}), if we redefine the attenuated post-selected Bloch vector (\ref{eq:attenuatedPostSelection}) to
\begin{equation}\label{eq:attenuatedPostSelectionNonMarkovian}
    \vec{f_I^\gamma}\left(t+\tau\right)=
    \begin{pmatrix}
        f_{Ix}\ \Gamma\\
        f_{Iy}\ \Gamma\\
        f_{Iz}\ \Gamma^2
    \end{pmatrix}.
\end{equation}

\bibliography{biblioWithDOI}






\end{document}